\documentclass[%
 reprint,
nofootinbib,
 amsmath,amssymb,
 aps,
]{revtex4-1}

\usepackage{graphicx}% Include figure files
\usepackage{dcolumn}% Align table columns on decimal point
\usepackage{bm}% bold math
\usepackage{physics}
\usepackage{slashed}
\usepackage{xcolor}
\usepackage{dsfont}
\usepackage[acronym]{glossaries}
\usepackage[colorlinks = true,linkcolor=blue,citecolor = blue]{hyperref}% add hypertext capabilities
\newif\ifSMver
\SMverfalse

\ifSMver
\newcommand{\mysec}[1]{{\em #1} ---}
\newcommand{\app}[1]{Sec.~\ref{#1} of the Supplemental Material}
\newcommand{\appendixsection}{%
\setcounter{equation}{0}
\setcounter{section}{0}
\setcounter{figure}{0}
\renewcommand{\theequation}{S\arabic{equation}}
\renewcommand{\thefigure}{S\arabic{figure}}
\onecolumngrid
\vspace*{.7cm}
\hrule
\vspace*{.04cm}
\hrule
\begin{center}
\vspace*{.4cm}
{\bf \large Supplemental Material}
\vspace*{.5cm}
\end{center}
\twocolumngrid
}
\else
\newcommand{\mysec}[1]{\section{#1}}
\newcommand{\app}[1]{App.~\ref{#1}}
\newcommand{\appendixsection}{\appendix}
\fi

\newcommand{\CME}{C_{\text{CME}}}
\newcommand{\chime}{\chi_{\text{CME}}}

\newacronym{hic}{HIC}{Heavy-ion collision}
\newacronym{qcd}{QCD}{Quantum Chromodynamics}
\newacronym{cme}{CME}{Chiral Magnetic Effect}

\begin{document}

\preprint{APS/123-QED}

\title{Localized Chiral Magnetic Effect in equilibrium QCD}

\author{Bastian B. Brandt}
%\email{brandt@physik.uni-bielefeld.de}
\author{Gergely Endr\H{o}di}
%\email{endrodi@physik.uni-bielefeld.de}
\author{Eduardo Garnacho-Velasco}
%\email{egarnacho@physik.uni-bielefeld.de}
\author{Gergely Mark\'{o}}
%\email{gmarko@physik.uni-bielefeld.de}
\author{A. Dean M. Valois}%
%\email{dvalois@physik.uni-bielefeld.de}
\affiliation{Fakult\"{a}t f\"{u}r Physik, Universit\"{a}t Bielefeld,\\
Universit\"{a}tsstra{\ss}e 25, 33615 Bielefeld, Germany}

\date{\today}% It is always \today, today,

\begin{abstract}
We study the impact of a non-uniform magnetic background field on the Chiral Magnetic Effect (CME) in equilibrium QCD using lattice simulations with 2+1 flavors of dynamical staggered quarks at
the physical point. We show that in the presence of a non-uniform magnetic field the CME manifests itself via a localized electromagnetic current density along the direction of the field, which integrates to zero over the
full volume. Our primary observable is the leading-order coefficient of the current in a chiral chemical potential expansion, which we compute for various lattice spacings and extrapolate to the continuum limit. Our findings demonstrate that, even though the global spatial average of the CME conductivity vanishes in equilibrium, steady currents still exist locally. Thus, spatially modulated magnetic fields provide a possible way of generating a non-trivial CME signal in equilibrium. 
\end{abstract}

\maketitle

%\tableofcontents

\mysec{Introduction}
\label{sec:intro}
\gls{hic} experiments provide an exceptional environment to investigate strongly interacting matter by exposing it to extreme conditions. A prime example of effects arising in this context are anomalous transport phenomena. These effects relate quantum anomalies with electromagnetic fields and vorticities, producing a series of non-dissipative currents which are the subject of extensive theoretical and experimental studies (see Ref.~\cite{Kharzeev:2024zzm} for a recent review). Being intimately related to anomalies, these effects effectively probe the topological nature of the vacuum of \gls{qcd}. 

The most celebrated among anomalous transport phenomena is the \gls{cme}: the generation of an electromagnetic current in a magnetized and chirally imbalanced system \cite{Fukushima:2008xe}. This effect is now understood as an out-of-equilibrium phenomenon, which has been linked with negative magnetoresistance in Dirac semimetals~\cite{Li:2014bha} and is actively sought for in heavy-ion collision experiments~\cite{STAR:2013ksd,STAR:2014uiw,STAR:2021mii}. In systems in thermal equilibrium, a global \gls{cme} current is absent. This can be understood as a direct consequence of the quantum-field-theoretical generalization of Bloch's theorem, which forbids global conserved currents to flow in equilibrium~\cite{Yamamoto:2015fxa}. On the quantum field theory level, regularization plays a crucial role in the absence of the \gls{cme} current~\cite{Hou:2011ze,Buividovich:2013hza,Zubkov:2016tcp,Buividovich:2024bmu,Horvath:2019dvl}. In lattice regularization for example, the use of a conserved vector current is of particular importance~\cite{Brandt:2024wlw}.

The vast majority of studies of anomalous transport effects, \gls{cme} in particular, have focused on homogeneous magnetic fields. However, the magnetic fields created in heavy-ion collisions are far from being uniform~\cite{Deng:2012pc}. 
These inhomogeneous fields have already been shown to have a sizeable effect on \gls{qcd} observables~\cite{Brandt:2023dir}. In this letter, we will use lattice \gls{qcd} simulations to investigate the impact of such magnetic fields on the chiral magnetic effect. Lattice simulations have been widely used to study anomalous transport effects \cite{Buividovich:2009wi, Yamamoto:2011gk, Yamamoto:2011ks, Buividovich:2013hza, Brandt:2024wlw, Buividovich:2024bmu, Buividovich:2020gnl, Puhr:2017ddx, Brandt:2023wgf}, as well as to investigate the role of non-uniform magnetic fields in \gls{qcd} \cite{Brandt:2023dir,Brandt:2024blb} (see the review~\cite{Endrodi:2024cqn}), providing an ideal framework to study the relation between these. The impact of weak inhomogeneities on the \gls{cme} has also been studied within the Wigner-Weyl formalism in Ref.~\cite{Banerjee:2021vvn}. 

 The chiral magnetic effect arises in general in the presence of background magnetic fields $B$ and a chiral imbalance, parameterized by a chiral chemical potential $\mu_5$. Besides the inhomogeneity of the magnetic field, it is realistic from a phenomenological point of view to consider the chiral imbalance to be non-uniform as well. In this letter, we will consider two scenarios including inhomogeneous magnetic fields: one with a uniform chiral chemical potential and one with $\mu_5$ exhibiting a similar inhomogeneity as the magnetic field itself.

This letter is organized as follows: In Sec.~\ref{sec:cme_non-uniform} we discuss the relevant observables that we compute to study the \gls{cme} as well as the details of the non-uniform magnetic background. This is followed by Sec.~\ref{sec:lattice_setup}, where we give the details of our simulation setup. Our results in full \gls{qcd} are presented in Sec.~\ref{sec:results_hom} for the homogeneous chiral imbalance, while in Sec.~\ref{sec:inhomogeneous_mu5}, we briefly explore the scenario with inhomogeneous $\mu_5$. Finally, we conclude in Sec.~\ref{sec:conclusions}. In a series of appendices, 
we provide an analytical calculation of our observable for free fermions, discuss the results with free staggered fermions, and give the details of our lattice implementations.

\mysec{CME and non-uniform magnetic fields}\label{sec:cme_non-uniform}
Throughout this letter, we consider \gls{qcd} in thermal equilibrium, in the presence of a non-uniform background magnetic field pointing in the third spatial direction. In particular, we choose an $x_1$-dependent profile of the form
\begin{equation}
    \vec{B}(x_1) = B\cosh^{-2}(x_1/\epsilon)\,\vec{e}_3\,,
\label{eq:inv_cosh2_profile}
\end{equation}
motivated by its analytical properties~\cite{Dunne:2004nc}.
The parameter $\epsilon$ sets the width of the field profile and is chosen as $\sim0.6$ fm in our \gls{qcd} simulations, in order to make contact with the \gls{hic} situation~\cite{Deng:2012pc}. Notice that the limit $\epsilon\to\infty$ corresponds to the homogeneous profile. The profile~\eqref{eq:inv_cosh2_profile} has already been used in \gls{qcd} models~\cite{Cao:2017gqs}, as well as on the lattice to study thermodynamics~\cite{Brandt:2023dir,Brandt:2024blb}.

The continuum electromagnetic current is defined as
\begin{equation}
j_{\nu}(x)=\sum_f \dfrac{q_f}{e} \, \bar\psi_f(x) \gamma_\nu \psi_f(x)\,,
\label{eq:veccurdef}
\end{equation}
where $f=u,d,s,\ldots$ labels the quark flavors, $q_f$ are the corresponding electric charges and $e$ is the elementary electric charge. Similarly, we consider the axial current
\begin{equation}
    j_{\nu5}(x) = \sum_f \, \bar\psi_f(x) \gamma_\nu \gamma_5  \psi_f(x) \,,
    \label{eq:axcurdef}
\end{equation}
associated with the total quark number, i.e.\ each quark flavor contributes with unit weight in it. The chiral chemical potential $\mu_5$, which parameterizes the chiral density, couples to the fourth component of Eq.~\eqref{eq:axcurdef}.
Below, we will also consider the currents averaged over a space-time slice, i.e.\ $J_\nu(x_1) \equiv T/L^2 \int \dd^4x' j_\nu(x'
)\delta(x_1-x_1')$ and similarly for $J_{\nu5}(x_1)$.

To study the chiral magnetic effect on the lattice, we follow a similar approach as in our previous work~\cite{Brandt:2024wlw}, except that the magnetic field is not assumed to be homogeneous but is given by Eq.~\eqref{eq:inv_cosh2_profile}. To make contact with the heavy-ion-collision-inspired setups described in Sec.~\ref{sec:intro}, we define the most general form of the \gls{cme} current in the presence of magnetic fields and chiral chemical potentials, where both depend on the $x_1$ coordinate. To linear order in $\mu_5$ and in $B$, the current parallel to the magnetic field reads,
\begin{equation}
    \label{eq:cme}
    \langle J_3(x_1)\rangle\!=\!\!\int\!\! \dd x_1' \dd x_1''\, \chime(x_1-x_1';x_1-x_1'')eB(x_1')\mu_5(x_1'')\,,
\end{equation}
involving the form factor $\chime$. In our simulations, which are performed at $\mu_5=0$~\footnote{Although simulations at non-zero chiral chemical potential are free of the sign problem, there are technical difficulties that arise when implementing $\mu_5$ in the staggered fermion formulation, which is the reason why we consider the first derive of the current with respect to $\mu_5$. See~\cite{Brandt:2024wlw} for a detailed discussion.}, it can be accessed via the first derivative, 
\begin{align}
    \label{eq:inhommu5_1}
    H(x_1,x''_1)&\equiv 
    \eval{\fdv{\langle J_{3}(x_1) \rangle}{\mu_5(x''_1)}}_{\mu_5=0} \\
    &=\int \dd x_1' \,\chime(x_1-x_1';x_1-x_1'')\,eB(x_1')\,, \nonumber
\end{align}
which describes the electromagnetic current generated at $x_1$ due to a weak chiral imbalance present at $x_1''$.
Note that $H(x_1,x_1'')$ depends on both arguments separately due to the breaking of translational invariance by the inhomogeneous magnetic field.

The response to a homogeneous chiral imbalance follows from replacing $\mu_5(x_1'')$ by $\mu_5$ in Eq.~\eqref{eq:cme}, resulting in the integral of $\chime$ over its second variable,
\begin{equation}
    \CME (x_1) = \int \dd x_1''\, \chime(x_1;x_1'')\,,
\end{equation}
which we refer to as the CME coefficient. For the homogeneous $\mu_5$ setup, Eq.~\eqref{eq:inhommu5_1} simplifies to
\begin{equation}
    G(x_1)
   \!\equiv\!
   \eval{\pdv{\langle J_{3}(x_1) \rangle}{\mu_5}}_{\mu_5=0}
   \!\!=\!\! \int\! \dd x_1'\, C_{\text{CME}}(x_1-x_1') \, e B(x_1')\,,
   \label{eq:j35der}
\end{equation}
which can also be constructed directly from Eq.~\eqref{eq:inhommu5_1} as $G(x_1)=\int \dd x_1''\, H(x_1,x_1'')$. Equivalently, Eq.~\eqref{eq:j35der} in Fourier space reads
\begin{equation}\label{eq:Cp}
   \widetilde{G}(q_1)=\widetilde{C}_{\rm CME}(q_1) \, e\widetilde{B}(q_1)\,.
\end{equation}

In the case of a homogeneous magnetic field, Eq.~\eqref{eq:j35der} trivially reduces to the global effect, parameterized by a single coefficient $\CME$, which coincides with the zero momentum limit of Eq.~\eqref{eq:Cp}, $\widetilde{C}_{\rm CME}(q_1=0)$. In Ref.~\cite{Brandt:2024wlw}, we showed that $\CME=0$ in full \gls{qcd}, in accordance with the discussion in Sec.~\ref{sec:intro} and, in particular, with Bloch's theorem. As we will show below, this picture is not changed by an inhomogeneous field. However, while Bloch's theorem forbids global currents to flow in equilibrium, it allows the appearance of local currents, and hence non-vanishing $G$ and $H$. Such local, $x_1$-dependent currents are in the focus of our attention here. Our main discussion in Sec.~\ref{sec:results_hom} revolves around $G$, while the more general behavior of $H$ is discussed in Sec.~\ref{sec:inhomogeneous_mu5}.

We note that for simplicity, we normalize our QCD results below by the overall proportionality factor $C_{\rm dof}=N_c\sum_f(q_f/e)^2$, where $N_c=3$ is the number of colors. In Apps.~\ref{app:analytical_cme} and \ref{app:results_free_fermions}, where we consider one colorless fermion flavor with charge $e$, this factor trivially reduces to $1$.

\mysec{Lattice setup}\label{sec:lattice_setup}
In our lattice \gls{qcd} simulations, we consider $2+1$ flavors of stout-smeared rooted staggered fermions with physical masses. In this formalism, the partition function $\mathcal{Z}$ can be written using the Euclidean path integral over the gluon links $U$, 
\begin{equation}
    \mathcal{Z}=\int \mathcal{D} U \exp[-\beta S_g] \,\prod_f \qty[ \det M_f(U,q_f,m_f)]^{1/4},
    \label{eq:partfunc}
\end{equation}
where the fermionic fields have already been integrated out to yield the fermion determinant, $\beta=6/g^2$ denotes the inverse gauge coupling and $m_f$ are the quark masses for each flavor $f=u,d,s$. In Eq.~\eqref{eq:partfunc}, $S_g$ is the gluonic action, for which we use a tree-level improved Symanzik action, and $M_f$ is the massive staggered Dirac operator, which contains the twice-stout-smeared links and the quark charges $q_u/2=-q_d=-q_s=e/3$. The quark masses are tuned to the physical point as a function of the lattice spacing $a$~\cite{Borsanyi:2010cj}. 

The simulations are performed on a four-dimensional lattice with $N_s$ spatial and $N_t$ temporal points. The physical spatial volume is given by $V=L^3=(aN_s)^3$ and the temperature by $T=(aN_t)^{-1}$.
At a fixed temperature, the continuum limit corresponds to $N_t\to\infty$. Note that the periodic spatial boundary conditions imply that the flux of the magnetic field is quantized. For our specific profile~\eqref{eq:inv_cosh2_profile}, this quantization condition takes the form~\cite{Brandt:2023dir},
\begin{equation}
eB = \frac{3\pi N_b}{\epsilon L\tanh(L/2\epsilon)}\,,\hspace{0.5cm}\text{where } N_b \in \mathbb{Z}\,.
\end{equation}
Furthermore, we note that our setup does not include dynamical photons, i.e.\ the magnetic field is treated as a classical background field.

The derivative~\eqref{eq:j35der} -- and the functional derivative~\eqref{eq:inhommu5_1} -- of the current with respect to $\mu_5$ results in a current-axial current correlator. We use the conserved (one-link) vector current and the anomalous axial (three-link) current in the staggered formulation~\cite{Durr:2013gp}. We stress that for staggered quarks, care has to be taken when evaluating the $\mu_5$ derivative of the current, for a tadpole term also appears. The exact form of the observables is given in \app{app:valence}, and a detailed discussion on the currents can be found in Ref.~\cite{Brandt:2024wlw}.

\begin{figure}[t]
\includegraphics[width=0.49\textwidth]{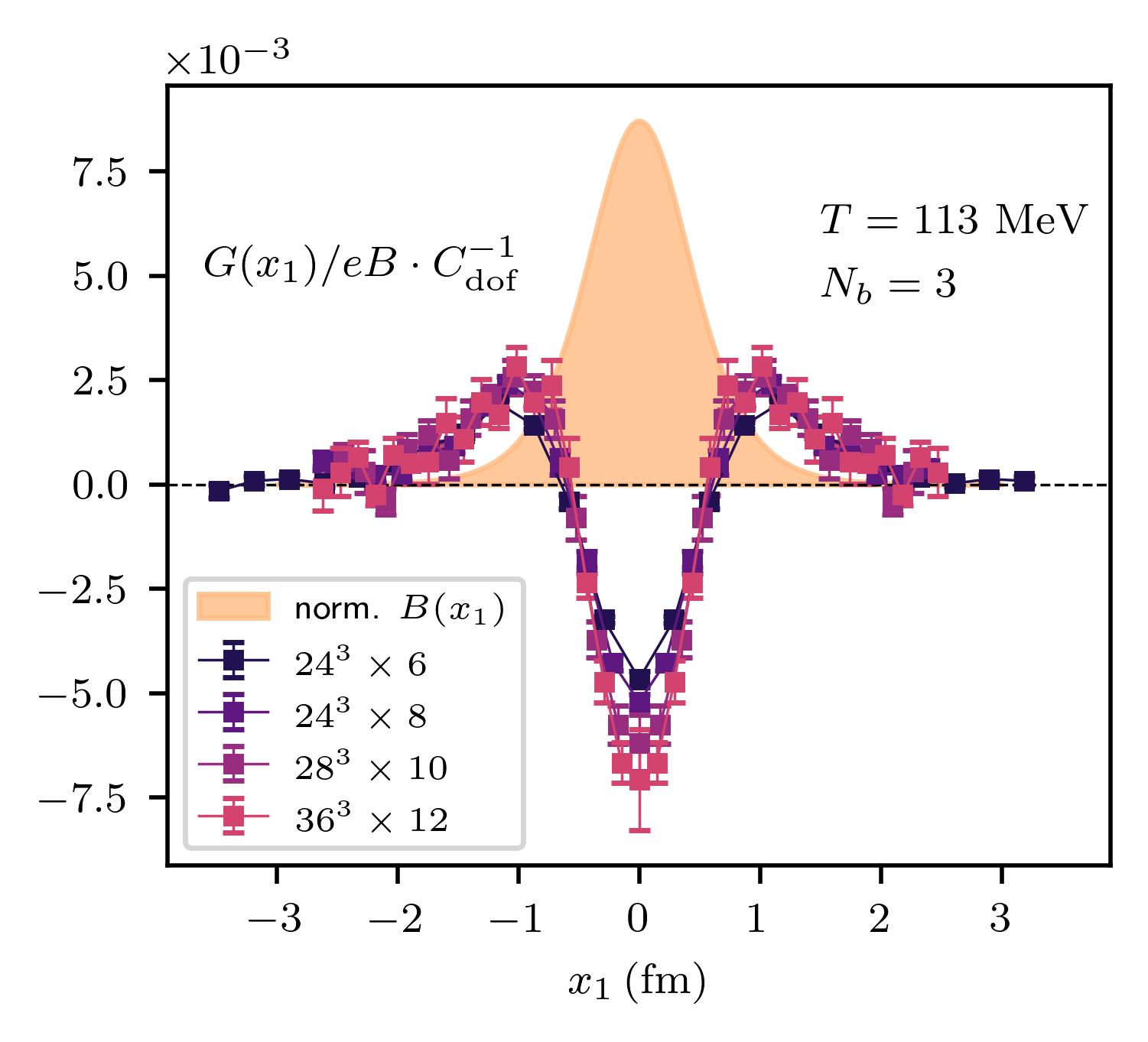}
\caption{Lattice data for the CME correlator in \gls{qcd} as a function of $x_1$ at $T = 113$ MeV and $N_b = 3$ for different lattice spacings. The connecting lines serve to guide the eye. For comparison, the shaded area depicts the magnetic field profile from Eq.~\eqref{eq:inv_cosh2_profile} in an arbitrary normalization. Notice that the different data sets are at slightly different physical $B$ values because of the change in physical volume.}
\label{fig:local_cme_corr_lattice}
\end{figure}

\mysec{Homogeneous chiral chemical potential}
\label{sec:results_hom}
In this section, we discuss the results for the $x_1$-dependence of the \gls{cme} correlator~\eqref{eq:j35der} for a homogeneous $\mu_5$. For non-interacting fermions, $G(x_1)$ can be calculated analytically, see \app{app:analytical_cme} for further details. As a cross-check of our lattice setup, we computed the correlator for free staggered fermions as well. 
The continuum-extrapolated lattice results were found to match the analytic formula perfectly, serving as a validation of our lattice setup. The details of the lattice calculation for free staggered fermions are also discussed in \app{app:results_free_fermions}.

After this cross-check,
we continue by turning on color interactions and analyzing the results in full \gls{qcd}. We take into account both connected and disconnected contributions to the CME correlator.
In Fig.~\ref{fig:local_cme_corr_lattice}, we show $G(x_1)$ as a function of $x_1$ for different lattice spacings and a weak magnetic field with profile width $\epsilon\approx0.6\textmd{ fm}$.
The figure shows that in this background, the \gls{cme} correlator develops a non-trivial spatial structure, with the current flowing in opposite directions in the center ($x_1\approx0$) and towards the edges ($|x_1|\gtrsim\epsilon$). While the physical magnetic fields slightly vary for fixed $N_b$ for the different ensembles, it is clear that the qualitative behavior persists across different lattice spacings. This behavior is similar to what we observed in the free case (see \app{app:results_free_fermions}). The current profile integrates to zero, implying that the global \gls{cme} conductivity vanishes in equilibrium \gls{qcd}, in agreement with our earlier findings~\cite{Brandt:2024wlw}.
We stress that our observable depends implicitly on the magnetic field profile.

\begin{figure}[t]
\includegraphics[width=0.49\textwidth]{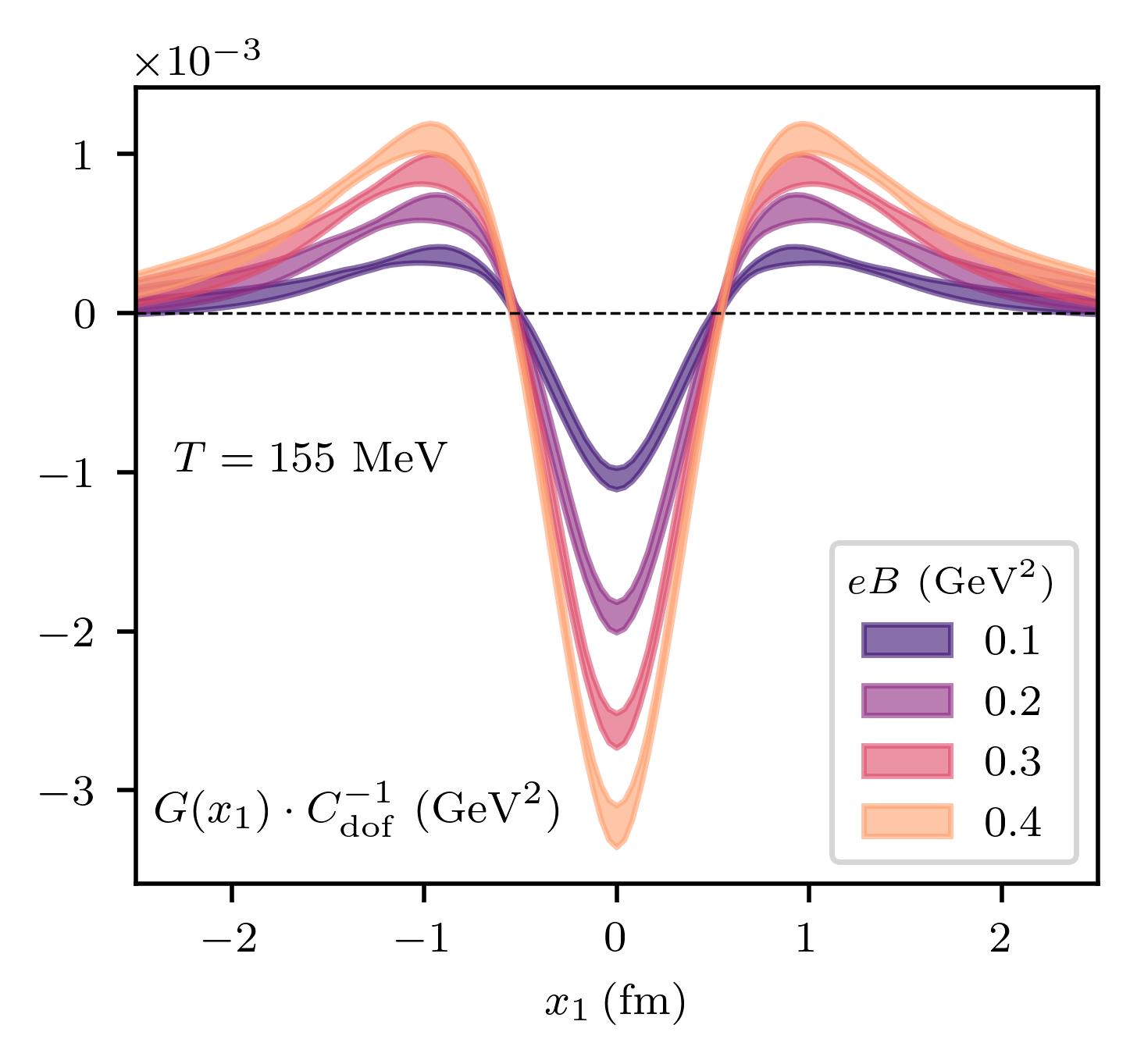}
\caption{Continuum limit of the CME correlator in \gls{qcd} as a function of $x_1$ at $T = 155$ MeV. The bands show this correlator at three different magnetic field strengths: $eB = 0.1, 0.2, 0.3$ and $0.4$ GeV$^2$. 
}
\label{fig:continuum_limit_cmex}
\end{figure}

\begin{figure}[t]
\includegraphics[width=0.49\textwidth]{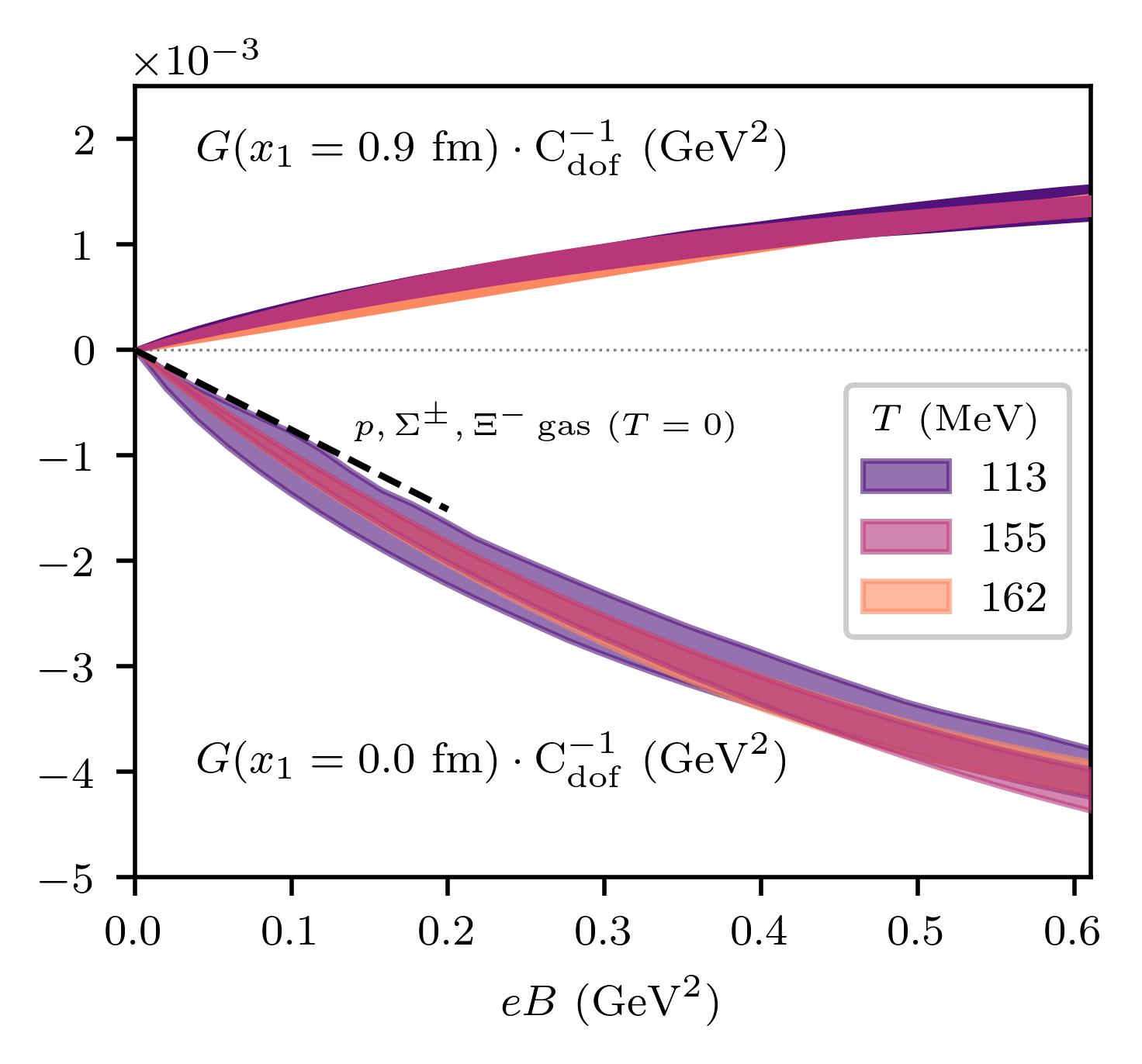}
\caption{CME correlator in the continuum limit at $x_1=0$ fm and at $x_1=0.9$ fm as a function of $eB$ for temperatures below, at, and above the crossover. The black dashed line corresponds to the $x_1=0$ value of the correlator obtained in a hadron resonance gas-type model, see App.~\ref{app:analytical_cme} for a detailed discussion. 
}
\label{fig:continuum_limit_cmex_B}
\end{figure}

Based on the results obtained on four different lattice spacings and different weak magnetic fields, we carry out the continuum extrapolation of $G(x_1)$, employing a multi-dimensional spline fit in $x_1$, $a$ and $eB$ with nodepoints generated via Monte Carlo, see App.~\ref{app:spline}. %~\cite{endrHodi2011multidimensional,Brandt:2022hwy,Brandt:2016zdy}.
In Fig.~\ref{fig:continuum_limit_cmex}, we show the so obtained continuum limit of the $x_1$-dependent CME correlator for various values of $eB$. 
Finally, in Fig.~\ref{fig:continuum_limit_cmex_B}, we show the CME correlator at the center ($x_1=0$) and near the edge ($x_1=0.9\textmd{ fm}$) as a function of the magnetic field. As one can see, the $B$-dependence is not affected significantly by the temperature. In fact, the slope at the origin is around $-1/(2\pi^2)\cdot 0.2$ below, at, and slightly above the transition temperature (for comparison, the expected out-of-equilibrium conductivity coefficient is $1/(2\pi^2)$~\cite{Fukushima:2008xe}). To better understand this finding, we calculated the same observable in a hadron resonance gas-type model that we discuss in App.~\ref{app:analytical_cme}, see also~\cite{Brandt:2023wgf}. The model -- consisting of non-interacting proton, $\Sigma^\pm$ and $\Xi^-$ baryon degrees of freedom -- reproduces the $T$-independence in the confined phase and predicts a nonzero value at $T=0$, also indicated in Fig.~\ref{fig:continuum_limit_cmex_B}.
Thus, the virtual baryon-antibaryon pairs of the vacuum are affected by $B$ (and $\mu_5$) even at zero temperature -- similarly to other observables like the pressure or the magnetization, see e.g.~\cite{Endrodi:2013cs}.

Another interesting aspect of $G(x_1)$ is that the effects of the magnetic field on the quark determinant in~\eqref{eq:partfunc} are found to be suppressed. This is analogous to what was found in the case of electric currents induced by inhomogeneous magnetic fields according to Amp\'ere's law in \gls{qcd}~\cite{Brandt:2024blb}. Negligible magnetic effects on the quark sea imply that the CME correlator might also be computed on the lattice using a computationally cheaper technique, the so-called valence approximation, where the configurations are generated at vanishing magnetic fields and the observable carries all the $B$-dependence. Here we did not rely on this approximation but discuss its details in \app{app:valence}.

\mysec{Beyond homogeneous chiral chemical potentials}
\label{sec:inhomogeneous_mu5}
Next, we consider the situation with an inhomogeneous chiral chemical potential $\mu_5(x_1')$. The induced current at point $x_1$ is now given by the correlator $H(x_1,x_1')$ from Eq.~\eqref{eq:inhommu5_1}.
This generalized correlator allows us to get a clearer picture of the role of $\mu_5$ in the thermodynamic system. In particular, if a parametrization of the chiral imbalance profile is available, it permits a convolution between this profile and the two-point function, yielding a more realistic one-dimensional picture on how the CME current might behave in experiments.

\begin{figure*}[!ht]
    \centering
\includegraphics[width=\textwidth]{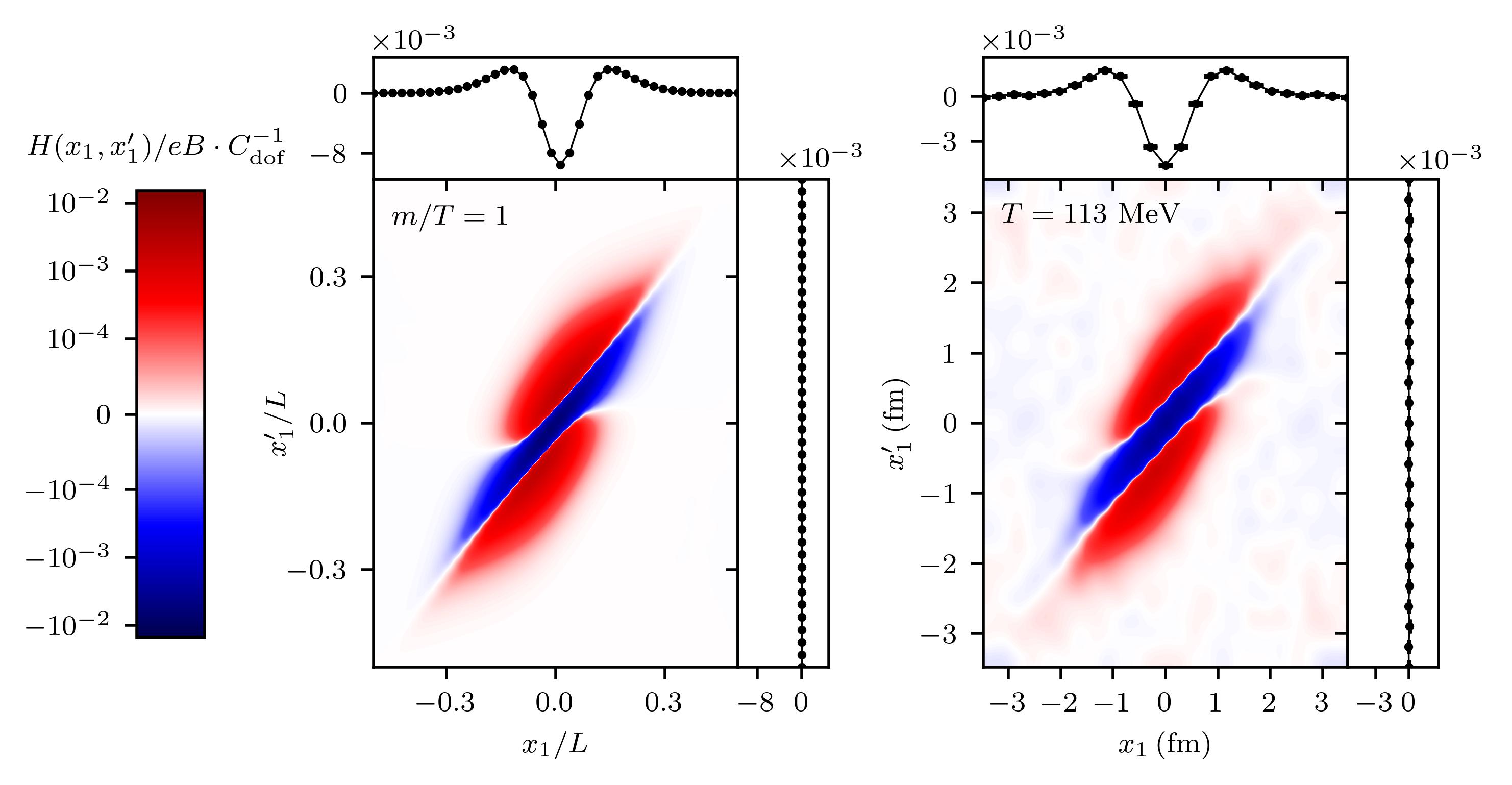}
    \caption{Left panel: heat plot of the correlator $H(x_1,x_1')$ normalized by $eB$ in the free case in the $x_1-x_1'$ plane for $m/T = 1$ on a $40^3\times10$ lattice. The color scheme is shown on a logarithmic scale for the absolute value of the observable from 0.015 to $10^{-4}$, and a linear scale from the latter to $0$. Red (blue) colors indicate the sign of the correlator. The projections on the top and right axes correspond to integrating over $x_1'$ and $x_1$, respectively. Right panel: the connected part of the same observable in \gls{qcd} at $T = 113$ MeV on a $24^3\times6$ lattice. To highlight the main features of the heat plot, the data was smoothed via a bicubic interpolation.}
    \label{fig:cmexx_heatmap}
\end{figure*}

In Fig.~\ref{fig:cmexx_heatmap}, we show the result for $H(x_1,x'_1)$ for a system of free fermions and for full \gls{qcd}. In the latter case, we neglected the disconnected contributions, which were merely found to enhance the noise, see \app{app:valence} for further details. The plots reveal more details about the local CME currents generated at different coordinates and the cancellations taking place in the global current. The latter can be understood by integrating over one of the two coordinates of $H(x_1,x'_1)$.

On the one hand, the sum over $x_1'$ corresponds to integrating out the spatial dependence of $J_{45}$, coming from $\mu_5$. As shown in the projection on the top axis of Fig.~\ref{fig:cmexx_heatmap}, this leads to a homogeneous $\mu_5$ effect and agrees with the profiles that we computed before, see Fig.~\ref{fig:local_cme_corr_lattice}. On the other hand, summing over the $x_1$ coordinate corresponds to the zero-momentum component of $J_3$ (projection on the right axis in Fig.~\ref{fig:cmexx_heatmap}), which vanishes due to Bloch's theorem.\footnote{We note that Bloch's theorem~\cite{Yamamoto:2015fxa} holds in the infinite volume limit. On a finite volume, we find that the sum of the profile in $x_1$ does give a nonzero signal, even though it is four orders of magnitude smaller than the one obtained by summing over $x_1'$. This signal is similar to the one found for the quark condensate in~\cite{Adhikari:2023fdl}, and vanishes in the thermodynamic limit.}

\mysec{Conclusions}\label{sec:conclusions}
In this letter, we studied the local spatial structure of the chiral magnetic effect in equilibrium \gls{qcd} with non-uniform magnetic background fields using first-principles lattice simulations. We found that in this setup -- in contrast to the situation with homogeneous magnetic fields -- nonzero electromagnetic currents flow in equilibrium. These CME currents are such that their spatial average vanishes, giving zero global current. 
Therefore, our results corroborate our earlier findings~\cite{Brandt:2024wlw} that the global CME vanishes in equilibrium, and demonstrate a novel behavior of the chiral magnetic currents in the presence of non-uniform magnetic fields.
From the theoretical point of view, it is intriguing to observe that the CME signal is realized as soon as the generalized Bloch theorem is circumvented by considering local currents instead of global ones.
Regarding experiments, such inhomogeneous magnetic fields and local currents are certainly more realistic for off-central heavy-ion collisions. 

Specifically, we analyzed the CME correlator, i.e.\ the current produced at $x$ due to a chiral imbalance at $x'$ and the inhomogeneous magnetic field.
This correlator can be convoluted with a given ansatz for the chiral chemical potential profile to predict the spatial dependence of the induced current.
Moreover, using the axial susceptibility $\chi_5$, the relationship between the chiral imbalance $n_5$ and the chiral chemical potential can also be constructed as $n_5=\chi_5 \mu_5$. Using our results for $\chi_5$ determined in Ref.~\cite{Brandt:2024wlw}, this allows one to predict CME signatures for a given fluctuation of chirality, leading to a more realistic description of the chiral magnetic current, albeit still in equilibrium. This, alongside our one-dimensional profiles, may be used to guide future phenomenological studies of the CME in the presence of non-uniform magnetic fields.  

Due to the highly inhomogeneous fields in \gls{hic}, our findings suggest that a signature of the CME may be hidden when looking at global observables, therefore constructing new ones sensitive to local signals, e.g.\ by implementing strict acceptance cuts in pseudo-rapidity or resolving higher harmonics in the azimuthal angle, might be necessary. Such signals could display the existence of topologically non-trivial fluctuations in the \gls{qcd} matter produced in relativistic collisions.

Finally, we stress that in this work we focused exclusively on the static CME, present in \gls{qcd} in equilibrium. In contrast, time-dependent responses, generated by a time-dependent chiral imbalance, give rise to the out-of-equilibrium CME, which is more challenging to study using Euclidean lattice simulations, as these necessitate analytic continuation of Euclidean correlators~\cite{Buividovich:2024bmu,Brandt:2025now}.

\begin{acknowledgments}
This work was funded by the DFG (Collaborative Research Center CRC-TR 211 ``Strong-interaction matter under
extreme conditions'' - project number 315477589 - TRR 211) and by the Helmholtz Graduate School for Hadron and Ion Research (HGS-HIRe for FAIR), as well as by STRONG-2020 ``The strong interaction at the frontier of knowledge: fundamental research and applications'' which received funding from the European Union's Horizon 2020 research and innovation programme under grant agreement No 824093. GE also acknowledges funding from the Hungarian National Research, Development and Innovation Office (Research Grant Hungary 150241) and the European Research Council (Consolidator Grant 101125637 CoStaMM).
The authors are grateful for enlightening discussions with Kenji Fukushima, Mikl\'os Horv\'ath, Xu-Guang Huang, Matthias Kaminski and Fuqiang Wang.
\end{acknowledgments}

\appendixsection
\renewcommand{\thesection}{\Alph{section}}

\section{Free fermions with Pauli-Villars regularization}\label{app:analytical_cme}
In this appendix, we calculate the \gls{cme} coefficient required to construct the current signal in an inhomogeneous background magnetic field for a single, colorless fermion flavor with mass $m$ and charge $e$. We carry out the calculation in Fourier space and obtain $\widetilde{C}_{\rm CME}(q_1)$ as used in Eq.~\eqref{eq:Cp}. In principle, the calculation outlined could also be used for the more general case, where the chiral chemical potential is also space dependent, but here we only concentrate on the momentum dependence corresponding to the inhomogeneity of the magnetic field.

We start by writing down the axial vector-vector-vector three-point correlator depending on two external momenta,
\begin{widetext}
\begin{align}
    \expval{j_{\mu 5}(-p-q)j_{\nu}(q)j_{\rho}(p)} \equiv \Gamma^{\rm AVV}_{\mu\nu\rho}(p+q,q,p)=&-i\sum_{s=0}^3 c_s\int_K\frac{{\rm\, Tr\,}\left[\gamma_\mu\gamma_5(\slashed{K}+m_s)\gamma_\nu(\slashed{K}+\slashed{q}+m_s)\gamma_\rho(\slashed{K}+\slashed{q}+\slashed{p}+m_s)\right]}{(K^2-m_s^2)((K+q)^2-m_s^2)((K+q+p)^2-m_s^2)}\nonumber \\
& + (\{\nu,q\}\leftrightarrow\{\rho,p\})\,,\label{eq:GAVV_def}
\end{align}
\end{widetext}
where we use Pauli-Villars (PV) regularization for QED following the textbook~\cite{Itzykson:1980rh}, see also Ref.~\cite{Horvath:2019dvl}. This involves the physical fermion $s=0$ with $c_0=1$ and $m^2_0=m^2$, as well as the regulator fields $s>0$ with $c_{1,2}=-c_3=-1$, $m^2_{1,2} = m^2 + \Lambda^2$ and $m_3^2=m^2 + 2\Lambda^2$. The continuum limit entails taking $\Lambda\to\infty$. We also used the notation $\int_K$ for a Matsubara summation and spatial integration over the loop momentum $K=(i\omega_n,\vec k\,)$, with $\omega_n$ being the fermionic Matsubara frequencies at temperature $T$. 

The three-point correlator $\Gamma^{\rm AVV}_{\mu\nu\rho}(p+q,q,p)$ is the one appearing in the $\mathrm{U_A}(1)$ anomaly: the famous formula can be recovered by contracting $\Gamma^{\rm AVV}_{\mu\nu\rho}$ with $p^\mu+q^\mu$. Here, however, a different combination is relevant: we need to set the external momentum of the axial leg to zero to ensure the homogeneity of $\mu_5$. In order to represent the equilibrium solution, we need to consider nonzero spatial momentum $\vec q$ with $q_0=0$. To further simplify the calculation, we restrict ourselves to the case where the current and the magnetic field point in the $x_3$ direction. We choose a gauge for the magnetic field where only $A_2\neq0$. Hence, for the $x_1$ dependence we want to model, only a non-vanishing $q_1$ is needed. All in all, for the momentum dependent coefficient appearing in Eq.~\eqref{eq:Cp}, we need to evaluate
\begin{align}
    \widetilde{C}_{\rm CME}(q_1)=\frac{1}{q_1}\Gamma_{023}^{\rm AVV}(0,q_1,-q_1)\,.
\end{align}
Evaluating the trace and taking the proper limits in Eq.~\eqref{eq:GAVV_def} yields
\begin{widetext}
\begin{align}
    \widetilde{C}_{\rm CME}(q_1)=\frac{8}{q_1}\sum_{s=0}^3 c_s T\sum_n\int\frac{\dd^3k}{(2\pi)^3}&\Bigg[\frac{2q_1m_s^2-2k_0^2q_1}{(K^2-m_s^2)^2((K+q)^2-m_s^2)}+\frac{k_1+q_1}{(K^2-m_s^2)((K+q)^2-m_s^2)}\Bigg]_{k_0=i\omega_n}\,.
\end{align}
The summation over Matsubara frequencies leads to the Fermi-Dirac distribution $n_F(x)=(\exp(x/T)+1)^{-1}$ and its derivative. After performing the angular integrals, we find
\begin{align}
     \widetilde{C}_{\rm CME}(q_1)=-\frac{1}{2\pi^2q_1}\sum_{s=0}^3 c_s\int_0^\infty \dd k\, k \left(\frac{m_s^2(1/2-n_F(E_{k,s}))}{E_{k,s}^3}-\frac{k^2}{E_{k,s}^2}n'_F(E_{k,s})\right)\log\frac{(2k-q_1)^2}{(2k+q_1)^2}\,,
\end{align}
where we introduced $E_{k,s}=\sqrt{k^2+m_s^2}$. The remaining integral over $k$ can be carried out for the PV regulator fields, that is $s>0$. The infinitely heavy fermions do not contribute to the $T$ dependence, while in the vacuum their contribution is equivalent to taking $q_1\to0$, since the integral only depends on $q_1/m_s$. 
This zero-momentum limit for the regulator fields produces $-1/(2\pi^2)$.
Our final formula for the momentum-dependent coefficient reads
\begin{align}
\label{eq:ccmeq_free}
     \widetilde{C}_{\rm CME}(q_1)=-\frac{1}{2\pi^2}-\frac{1}{2\pi^2q_1}\int_0^\infty \dd k \,k \left(\frac{m^2(1/2-n_F(E_k))}{E_k^3}-\frac{k^2}{E_k^2}n'_F(E_k)\right)\log\frac{(2k-q_1)^2}{(2k+q_1)^2}\,,
\end{align}
\end{widetext}
with the physical energy $E_k=E_{k,0}$, which agrees with the results of Ref.~\cite{Hou:2011ze} obtained in a slightly different way. The final integral has to be performed numerically. We note that the homogeneous magnetic field limit, that is $q_1\to0$, results in a vanishing coefficient through a double cancellation: the nonzero temperature part is zero separately for the physical fermion, while the vacuum term cancels due to the contribution of the PV fields. This confirms our previous findings regarding the vanishing of the CME current in equilibrium with homogeneous magnetic fields~\cite{Brandt:2024wlw}.

The inverse Fourier transform of Eq.~\eqref{eq:Cp}, using $\widetilde{C}_{\rm CME}(q_1)$ from Eq.~\eqref{eq:ccmeq_free} and the specific magnetic profile, yields the CME correlator $G(x_1)$. This is shown by the black dashed line in Fig.~\ref{fig:CLfree} below.

One interesting observation is that the PV fields contribute only a constant shift to $\widetilde{C}_{\rm CME}(q_1)$. In other words, their contribution is a Dirac $\delta$ in coordinate space. This explains the shape of $G(x_1)$ in Fig.~\ref{fig:CLfree}, where one can recognize the sum of a sharp negative peak and a broader positive one. The former is the contribution of the regulator fields, for which the contact nature of $C_{\rm CME}(x_1)\propto\delta(x_1)$ results in a current profile directly proportional to $eB(x_1)$. In turn, for the 
physical fermion, the non-trivial $q_1$ dependence results in a smeared reaction to the magnetic profile in coordinate space.

\begin{figure}[!ht]
\includegraphics[width=0.49\textwidth]{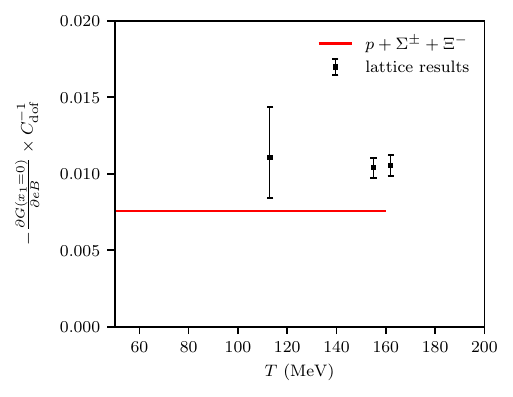}
\caption{Prediction of the baryon gas model for the negative slope of the correlator with respect to the magnetic field at $x_1=0$, compared to the full QCD lattice results.}
\label{fig:hrg}
\end{figure}

The above calculation also allows us to build a naive low temperature model consisting of non-interacting spin-$1/2$ charged baryon degrees of freedom. Including independent contributions for the protons as well as the $\Sigma^{\pm}$ and $\Xi^-$ baryons (masses taken from the PDG \cite{ParticleDataGroup:2024cfk}) yields an almost temperature independent result up to $T\approx160\textmd{ MeV}$. In Fig.~\ref{fig:hrg} we display the lattice results for the negative of the slope $\partial G(x_1=0)/\partial (eB)$ at the center, and compare it to the baryon gas model. We find that the two are consistent within two standard deviations of the lattice data.

Thus, contrary to standard $B=0$ thermodynamic observables, our correlator does not vanish at $T=0$. As mentioned in the main text, this is because virtual particle-antiparticle pairs that exist in the quantum vacuum are sensitive to the magnetic field (and the chiral chemical potential) and can contribute to $G(x_1)$. In this sense the correlator is similar to the magnetization, which is also nonzero at $B>0$ and $T=0$, owing to the spin and angular momentum carried by virtual particle pairs, see e.g.~\cite{Endrodi:2013cs}.

\section{Free fermions on the lattice}\label{app:results_free_fermions}
In this appendix we turn to the lattice discretization of Eq.~\eqref{eq:j35der} for non-interacting fermions. We again consider one colorless quark flavor with charge $e$ and mass $m$. We will cross-check our lattice approach against the Pauli-Villars regularization discussed in App.~\ref{app:analytical_cme}. The results for free fermions also reveal information about the high temperature limit of \gls{qcd}, where it can be described in terms of a gas of free massless fermions.

For the lattice calculation, we use the exact eigensystem of the staggered Dirac operator, reconstructing the required matrix elements to calculate $G(x_1)$. This approach has the advantage of not relying on stochastic estimators for the traces, yielding an exact result for the $x_1$ dependence of the operator. For more details on the exact diagonalization, see Ref.~\cite{Brandt:2024wlw}.   

\begin{figure}[t]
\includegraphics[width=0.49\textwidth]{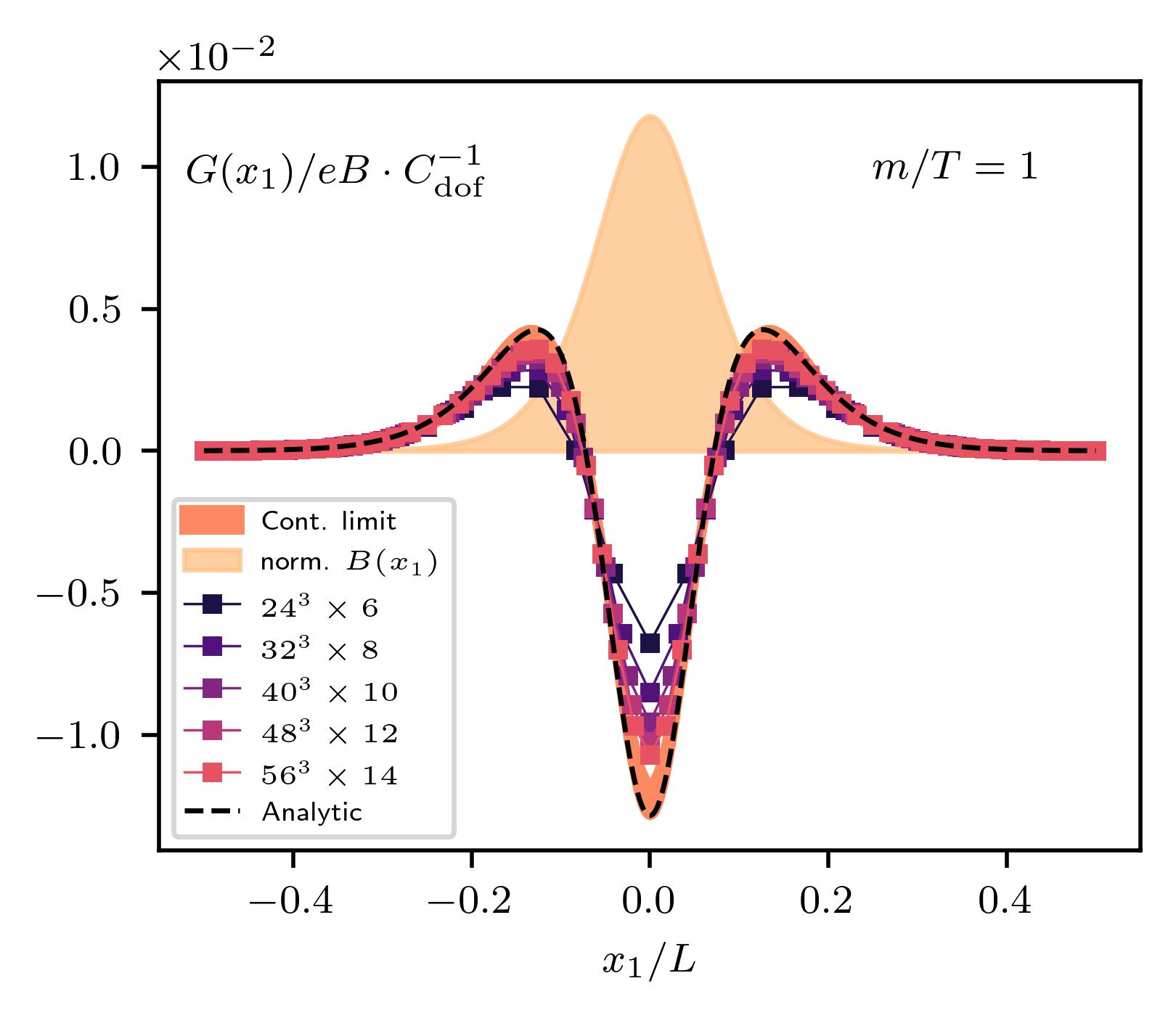}
\caption{Lattice data and continuum extrapolation of the CME correlator with $eB/T^2=14.14$ and $\epsilon T=1/3$, normalized by the magnetic field, for free fermions. The analytical result is recovered in the continuum limit, confirming the validity of
our setup. For comparison, the shaded area depicts the magnetic field profile in arbitrary normalization.}
\label{fig:CLfree}
\end{figure}

Fig.~\ref{fig:CLfree} shows an example of the continuum extrapolation of the \gls{cme} correlator for $m/T=1$, $N_b=2$ and aspect ratio $LT=4$. The latter choice was found to be sufficiently close to the thermodynamic limit so that finite volume effects are negligible. The continuum limit agrees with the analytical calculation, validating our lattice implementation. The $x_1$-integral of $G(x_1)$ is found to vanish -- in other words, $\widetilde{G}(q_1=0)=0$, confirming that no global CME current flows in equilibrium.

\begin{figure}[t]
\includegraphics[width=0.49\textwidth]{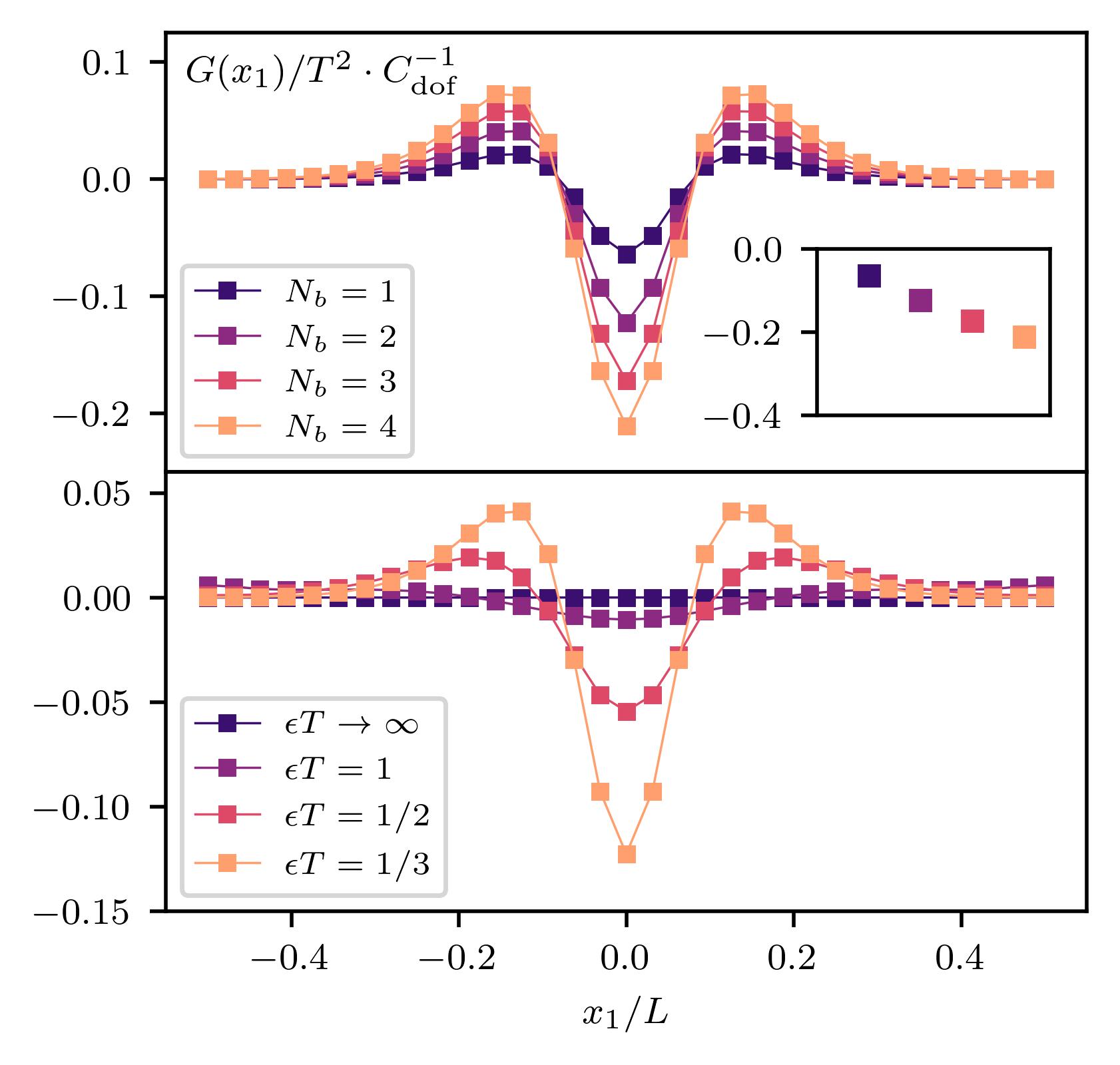}
\caption{ Top panel: CME correlator for free fermions as a function of $x_1/L$ for different values of $N_b$. The results correspond to a $32^3\times8$ lattice with $m/T=1$ and $\epsilon T=1/3$. The inset shows the magnetic field dependence of the central point. Bottom plot: $\epsilon$ dependence of the CME correlator, calculated on a $32^3\times8$ lattice with $m/T=1$ and $N_b=2$. Notice that the current vanishes in the limit of homogeneous magnetic fields, i.e.\ $\epsilon\to\infty$.}
\label{fig:epsnbdepfree}
\end{figure}

Next, we discuss how the \gls{cme} correlator is affected by the details of the magnetic profile.
In Fig.~\ref{fig:epsnbdepfree}, we show the impact of changing the magnitude (top panel) and the profile width (bottom panel) of the magnetic field. The middle point ($x_1=0$) scales linearly for weak $B$, a behavior that is found to persist in \gls{qcd} as well (see Fig.~3 in the main text). Increasing $\epsilon$, the shape of the \gls{cme} correlator flattens, until a homogeneous $B$ field is reached for $\epsilon\rightarrow \infty$. In this limit, we see that the correlator vanishes identically for every $x_1$, confirming our earlier findings~\cite{Brandt:2024wlw}.

\begin{figure}[!ht]
\includegraphics[width=0.49\textwidth]{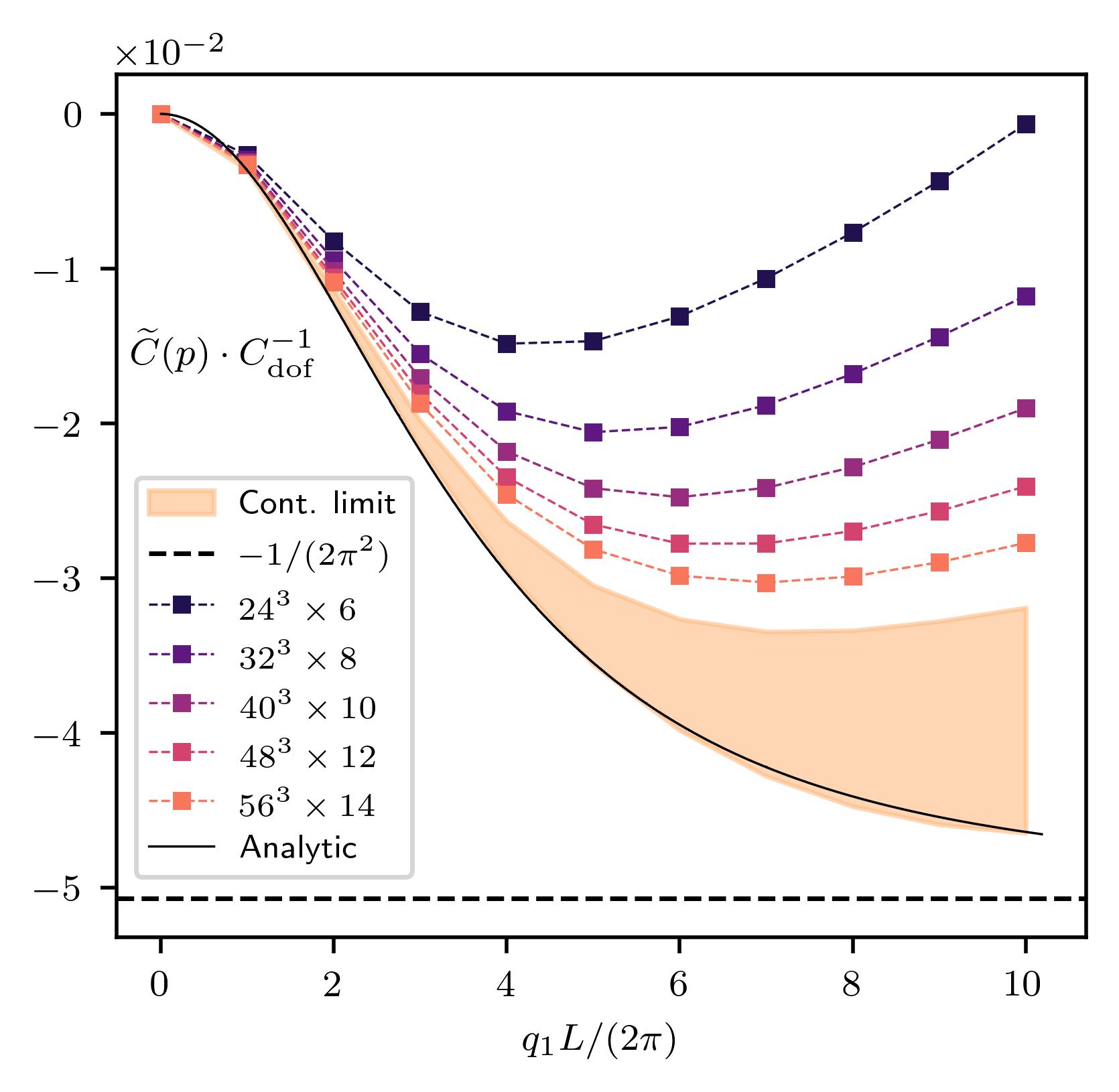}
\caption{Fourier transform of the CME coefficient as a function of the momentum for different lattice sizes. The solid line represents the analytical result. In the infinite-momentum limit, the result converges to $-1/(2\pi^2)$, which arises purely from the Pauli-Villars regulator fields in the analytic calculation. 
}
\label{fig:Cpfree}
\end{figure}

The vanishing of the global \gls{cme} can be also understood in momentum space. In Fig.~\ref{fig:Cpfree}, we show the Fourier transform of the \gls{cme} coefficient $\widetilde{C}_{\rm CME}(q_1)$. In this plot, the role of the regulator is most transparent. As we have seen in App.~\ref{app:analytical_cme}, the Pauli-Villars regulator fields contribute $-1/(2\pi^2)$, and this value is approached in the infinite momentum limit. Together with this contribution, the total \gls{cme} coefficient vanishes at zero momentum. Notice that the higher momentum components are increasingly more difficult to extract on the lattice, since large momenta cannot be resolved at finite lattice spacing. To determine the continuum limit, we fitted the data by including lattice artifacts up to quartic order in the lattice spacing and excluding the coarsest lattice.
The systematic error was estimated by performing similar fits, considering $\mathcal{O}(a^2)$ lattice artifacts and including all data points. The so defined error is shown by the yellow band in Fig.~\ref{fig:Cpfree}. It is found to overlap with the analytical result, demonstrating again that finite volume effects in the lattice results are small.

\section{Lattice implementations and valence approximation}\label{app:valence}
In this appendix, we examine the valence and sea contributions to the impact of the magnetic field in the CME and introduce the valence approximation. Moreover, we specify the details of our implementation of the CME correlators introduced in the main text.

In the expectation value of any fermionic operator $A$ with respect to the rooted staggered partition function, the dependence on the magnetic field enters in two distinct ways: in the operator (valence contribution) and in the fermion determinant (sea contribution),
\begin{equation}
\expval{A}=\int \mathcal{D}U\frac{e^{-\beta S_g}}{\mathcal{Z}(B)}\prod_f\,[\det M_f(B)]^{1/4}\,A(B)\,.
\end{equation}
The $B$ dependence in the operator appears via quark propagators $M_f^{-1}(B)$. Thus, in a perturbative language, the valence effect arises due to the coupling of the magnetic field to the valence quark propagator, while the sea effect is due to the coupling to virtual quark loops.
More specifically, the valence contribution reads
\begin{equation}
\expval{A}^{\rm val}=\int \mathcal{D}U\frac{e^{-\beta S_g}}{\mathcal{Z}(B=0)}\prod_f\,[\det M_f(B=0)]^{1/4}\,A(B)\,.
\label{eq:Avaldef}
\end{equation}
Conversely, the sea contribution corresponds to setting $B=0$ in the operator, but keeping it in the determinants,
\begin{equation}
\expval{A}^{\rm sea}=\int \mathcal{D}U\frac{e^{-\beta S_g}}{\mathcal{Z}(B)}\prod_f\,[\det M_f(B)]^{1/4}\,A(B=0)\,.
\end{equation}

For weak magnetic fields, expectation values $\expval{A}$ can be approximately decomposed into their valence $\expval{A}^{\rm val}$ and sea $\expval{A}^{\rm sea}$ contributions. For expectation values that are odd in the magnetic field -- like the CME current or the CME correlators -- this weak-field leading-order additivity relation takes the form~\cite{Brandt:2024blb},
\begin{equation}
\expval{A}=\expval{A}^{\rm val} + \expval{A}^{\rm sea} + \mathcal{O}(B^3)\,.
\end{equation}
We note that for expectation values even in $B$, a similar decomposition can also be derived~\cite{Brandt:2024blb}.
Often one finds empirically that the valence contribution dominates this sum so that $\expval{A}\approx \expval{A}^{\rm val}$ at leading order. This is the valence approximation, which we investigate below for the CME correlator as observable.

To do so, we give a detailed definition of our observables: the current and the CME correlators.
In the staggered formalism, the electric current expectation value reads
\begin{equation}
\expval{J_{\nu}(x_1)} =
 \frac{T}{4L^2}\sum_f \frac{q_f}{e}\left\langle \Tr \left(\mathcal{P}_{x_1}\Gamma^f_{\nu}M_f^{-1}(B)\right)  \right\rangle\,,
 \label{eq:jnustagg}
\end{equation}
where the trace entails sums over color and space-time coordinates and $\mathcal{P}_{x_1}$ denotes a projector to the slice of the lattice, where the first spatial coordinate equals $x_1$. In Eq.~\eqref{eq:jnustagg}, the staggered Dirac matrices $\Gamma_\nu^f$ appear. For their explicit form in the presence of $B$ and $\mu_5$, see Ref.~\cite{Brandt:2024wlw}. We will also need the staggered equivalents of the $\gamma_\nu\gamma_5$ matrices~\cite{Durr:2013gp},
\begin{equation}
\Gamma^f_{\nu 5} \equiv \frac{1}{3!}\sum_{\rho\alpha\beta}\epsilon_{\nu\rho\alpha\beta}\Gamma^f_{\rho}\Gamma^f_{\alpha}\Gamma^f_{\beta}\,.
\end{equation}
Its $\nu=4$ component couples to $\mu_5$ in the Dirac operator in an exponential form~\cite{Brandt:2024wlw}.

To obtain the correlator $H(x_1,x''_1)$ given in Eq.~\eqref{eq:inhommu5_1}, we need to perform the functional derivative of~\eqref{eq:jnustagg} with respect to $\mu_5(x_1')$. The chiral chemical potential appears in $M_f^{-1}$, in the determinants under the expectation value, in the normalization $\mathcal{Z}(B)$, as well as in $\Gamma_\nu^f$~\cite{Brandt:2024wlw}. Altogether, we arrive at
\begin{widetext}
\begin{align}
H(x_1,x_1^{\prime}) \equiv \frac{T}{4L^2}&\sum_f\left[\frac{1}{4}\sum_{f^{\prime}}\right.q_f\expval{\Tr[\mathcal{P}_{x_1}\Gamma^{f}_3M^{-1}_f(B)]\Tr[\mathcal{P}_{x_1^{\prime}}\Gamma^{f^{\prime}}_{45}M^{-1}_{f^{\prime}}(B)]}_{\!\!B}\nonumber\\
-&q_f\left.\expval{\Tr[\mathcal{P}_{x_1}\Gamma^{f}_3M^{-1}_f(B)\mathcal{P}_{x_1^{\prime}}\Gamma^f_{45}M^{-1}_f(B)]}_{\!\!B} + q_f\expval{\Tr\qty[\mathcal{P}_{x_1}\fdv{\Gamma^f_3}{\mu_5(x_1^{\prime})}M^{-1}_f(B)]}_{\!\!B}\,\right]\,.
\label{eq:full_H}
\end{align}
\end{widetext}
Here we used that $\expval{J_3(x_1)}$ vanishes at $\mu_5=0$, leaving the three terms on the right hand side, which we refer to as disconnected, connected and tadpole terms, respectively.
For clarity, in Eq.~\eqref{eq:full_H} we indicated that the expectation values are to be evaluated at nonzero $B$. 
The so calculated observable is shown in Fig.~4 of the main text, except that in the QCD case we ignored the disconnected contribution to it, which was merely found to dominate the statistical errors (see below).

Summing over the $x_1^{\prime}$ coordinate (i.e.\ the coordinate corresponding to the $\mu_5$ insertion) in~\eqref{eq:full_H} yields the correlator $G(x_1)$, in accordance with Eq.~\eqref{eq:j35der}.
The valence approximations, $H^{\rm val}(x_1,x_1')$ and $G^{\rm val}(x_1)$ of these correlators are obtained with the same operators evaluated at $B=0$, i.e.\ using the $B=0$ ensemble of gauge configurations, just like in the general case~\eqref{eq:Avaldef}. Moreover, we also consider the correlator without the disconnected term (evaluated at $B>0$), i.e.\ just the second and third terms in Eq.~\eqref{eq:full_H}, and $G^{\rm conn+tadp}$ calculated from it.

\begin{figure}[t]
\includegraphics[width=0.49\textwidth]{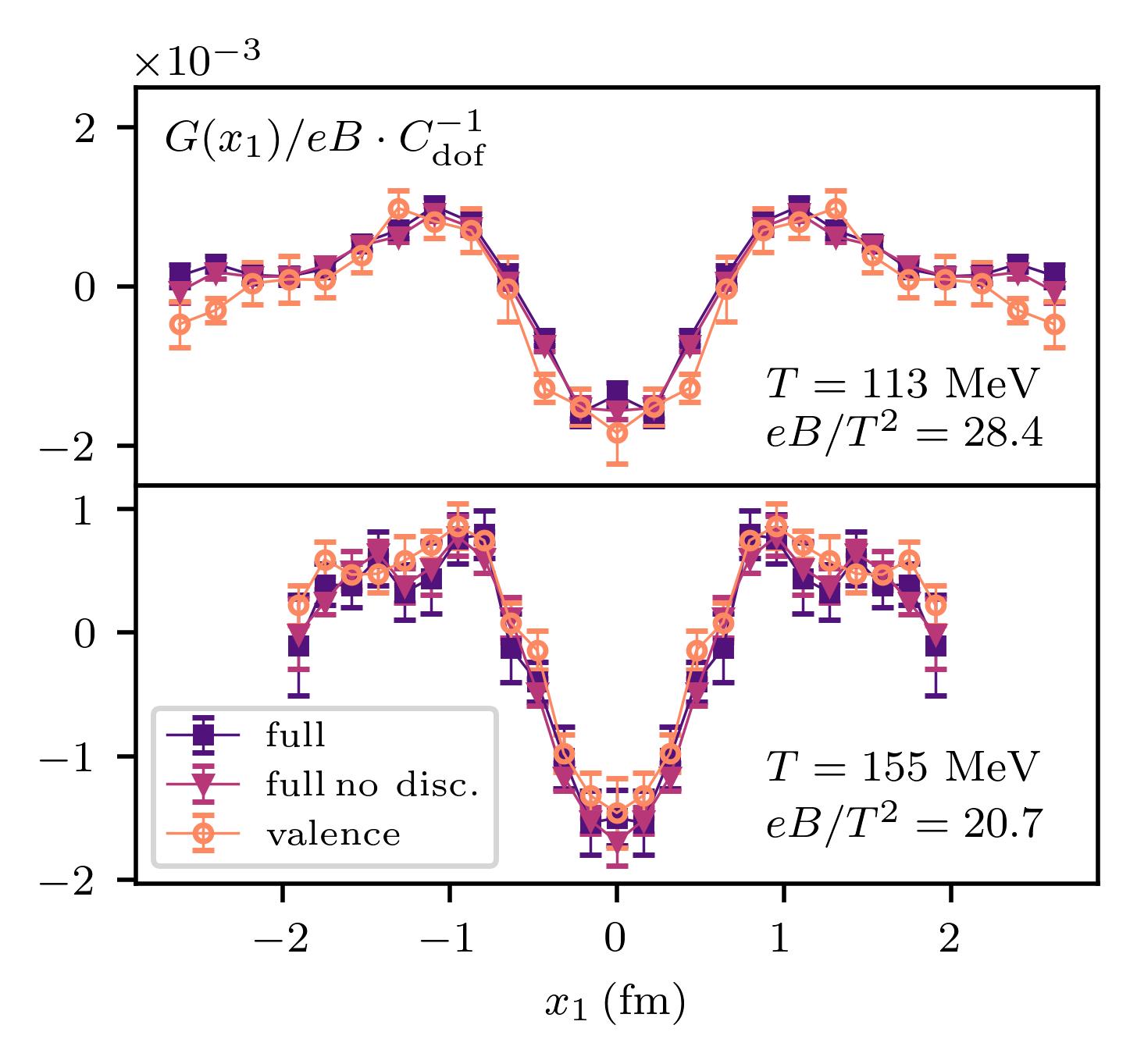}
\caption{Comparison between the full correlator $G(x_1)$, its valence contribution $G^{\rm val}(x_1)$ and the full correlator without disconnected contributions, $G^{\rm conn+tadp}(x_1)$, for a given magnetic field strength at low $T$ (upper plot) and at the crossover temperature (lower plot). The correlators are normalized by the magnetic field.}
\label{fig:valQCD}
\end{figure}

In Fig.~\ref{fig:valQCD} we compare $G(x_1)$, $G^{\rm val}(x_1)$ and $G^{\rm conn+tadp}(x_1)$ in QCD for two different temperatures and magnetic fields.
In both cases we observe a remarkable agreement among all three quantities, which we found to hold for all other simulation points as well. Interestingly, we observed the valence approximation to be valid even in the strong-field regime. In fact, we verified the approximate equality between the full and valence observables for magnetic fields as strong as $eB/T^2 \approx 83$ at $T = 113$ MeV, and $eB/T^2 \approx 114$ at $T = 155$ MeV. 

Altogether, these findings demonstrate that the sea effect is strongly suppressed compared to the valence one in this observable and, moreover, the disconnected terms -- contributing the most to the noise -- are also negligible. Since the valence approximation significantly reduces the computational costs, this approximation is suitable for studies involving more expensive fermion actions, such as Wilson, domain-wall, and overlap fermions, in the presence of background magnetic fields.

\section{Multidimensional continuum limit}\label{app:spline}
In this section, we give the details of our continuum limit approach based on~\cite{endrHodi2011multidimensional,Brandt:2022hwy,Brandt:2016zdy}, which we used to obtain the results for the CME correlator $G$ in Figs.~2 and~3 from the main text. In Table~\ref{tab:parameters}, we provide the details of the ensembles used in our dynamical QCD simulations at the physical point.

\begin{table}[b]
\caption{\label{tab:parameters}%
Ensembles used in the QCD simulations to obtain the continuum limits presented in the main text.
}
\renewcommand{\arraystretch}{1.2}
\begin{ruledtabular}
\begin{tabular}{cccc}
$T$ [MeV] &$N^3_s\times N_t$&$\beta$ &$N_b$\\
\colrule
\phantom{133} & $16^3\times6$ & 3.450  & 1,2,3,4\\
\phantom{133} & $24^3\times6$& 3.450  & 1,2,3,4\\
113 & $24^3\times8$& 3.546  & 1,2,3,4,6,8,10\\
\phantom{133} & $28^3\times10$& 3.623 & 1,2,3,4,6,8,10\\
\phantom{133} & $32^3\times12$ & 3.690 & 1,2,3,4,6,8,10\\
\colrule
\phantom{133} & $16^3\times6$ & 3.555  & 1,2,3,4\\
\phantom{133} & $24^3\times6$& 3.555  & 1,2,3,4\\
155 & $24^3\times8$& 3.657  & 1,2,3,4,6,8,10\\
\phantom{133} & $28^3\times10$& 3.742 & 1,2,3,4,6,8,10\\
\phantom{133} & $32^3\times12$ & 3.817 & 1,2,3,4,6,8,10\\
\colrule
\phantom{133} & $16^3\times6$ & 3.570 & 1,2,3,4,6,8\\
162 & $24^3\times8$& 3.673  & 1,2,3,4,6,8,10\\
\phantom{133} & $28^3\times10$& 3.759 & 1,2,3,4,6,8,10\\
\end{tabular}
\end{ruledtabular}
\end{table}

We use a bivariate quartic spline with lattice spacing dependent coefficients to obtain an interpolation of our data over the $x_1-eB$ plane at nonzero lattice spacings and towards the continuum limit. Each temperature is treated separately, in the following way. First of all, we exploit the parity symmetry $G(x_1) = G(-x_1)$ of the correlator, which enhances our statistics, as well as allows the interpolation to be restricted to the range $x_1\in[-L/2,0]$. We define a usual spline function on %(this we will come back to later how) 
a grid of node points, %, which do not coincide with $x_1,\ eB$ pairs where data is available [[except at the boundaries in $x_1$?]], 
and fit its value at each node by minimizing the usual sum of weighted squared residuals. Specifically, the spline value at each point is determined by two parameters: a constant and a term linear in $a^2$. This way, we account for the $a$-dependence in the fits and hence allow the combined use of data with different lattice spacings, assuming that all datasets are in the proper scaling regime. %Our splines are contiuous with continuous first and second derivatives in both direction, and fulfill an extra physical condition, namely having vanishing first $x_1$-derivatives at $x_1=0$ due to the symmetry mentioned earlier. 

\begin{figure}[t]
    \centering
    \includegraphics[width=0.49\textwidth]{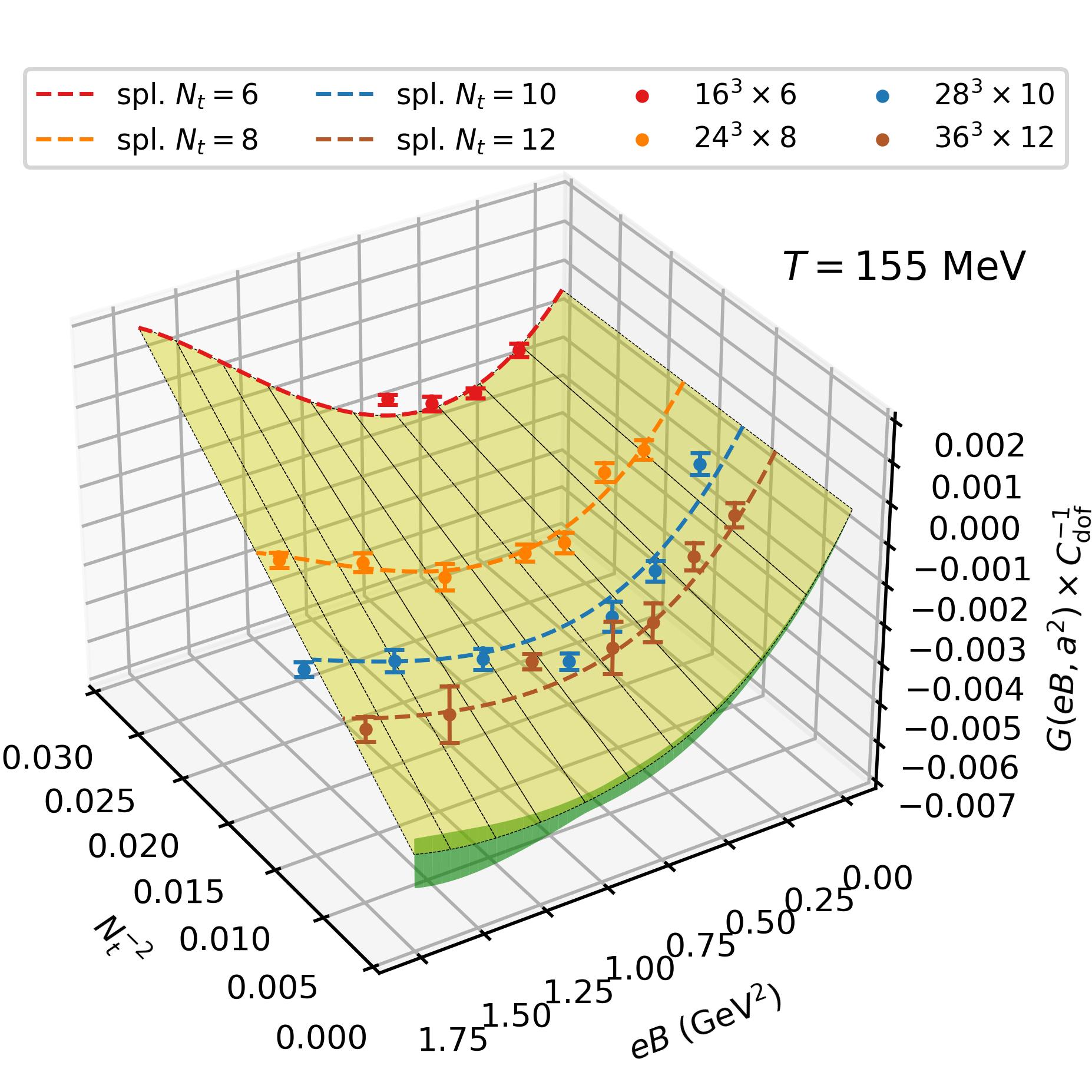}
    \caption{Lattice data for the CME correlator in the $eB-a^2$ plane at the point $x_1=0$ and $T=155$ MeV for different values of $N_t$. The green band corresponds to the continuum limit, whereas the dashed lines are the splines at each value of $N_t$.}
    \label{fig:continuum_limit_ab_plane}
\end{figure}

The fit procedure takes statistical fluctuations into account using the jackknife method, and at the same time smoothens out local outliers. To account for possible systematic errors arising from the choice of node points, we utilize a Monte Carlo routine to generate dozens of different grids with different number and placement of the node points. Our final interpolating function is then combined from each of the Monte Carlo samples based on the Akaike information criterion~\cite{Akaike:1974vps}, which takes into account the goodness of the fit as well as the number of free parameters. In this fashion, we create a model-independent interpolation of the data, in addition to carrying out the continuum limit.

\begin{figure}[t]
    \includegraphics[width=0.49\textwidth]{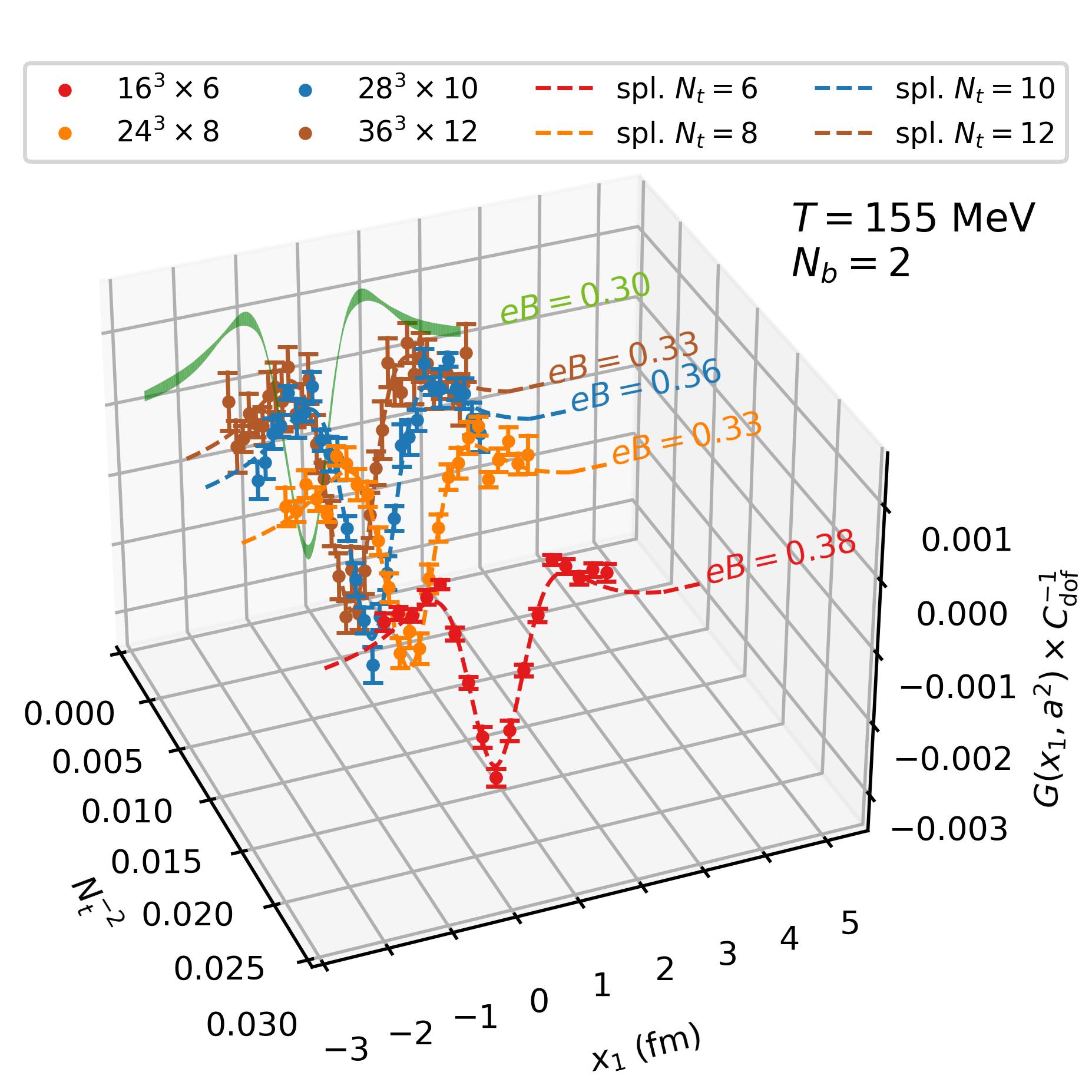}
    \caption{Lattice data for the CME correlator in the $x_1-a^2$ plane at $N_b=2$ and $T=155$ MeV for different $N_t$. The green band corresponds to the continuum limit at $eB = 0.3\text{ GeV}^2$, whereas the dashed lines are the splines at each value of $N_t$. Due to the slightly different volumes, the value of $N_b=2$ corresponds to slightly different values of $eB$ for some lattices, indicated in the plot.}
    \label{fig:continuum_limit_ax_plane}
\end{figure}

In Fig.~\ref{fig:continuum_limit_ab_plane}, we show a two-dimensional slice of the interpolating surface at the point $x_1=0$ in the $eB-a^2$ plane alongside the corresponding lattice data.
Notice that we excluded the lattice data for $N_t=6$ beyond $N_b=4$, in order to remain in the scaling region. %due to the fact that lattice artifacts are expected to be large there. 
The surface shows the approach towards the continuum limit, $N_t^{-2}\to0$.
The green band of Fig.~\ref{fig:continuum_limit_ab_plane} is also shown up to $eB\lesssim0.6\text{ GeV}^2$ in %Fig.~\ref{fig:continuum_limit_cmex_B}
Fig.~2 from the main text. In turn, in Fig.~\ref{fig:continuum_limit_ax_plane}, we visualize the same fit on a different slice, with the lattice data in the $x_1-a^2$ plane. Notice that here the individual lattice ensembles enter the combined fit at slightly different magnetic field values.

To summarize, our method allows for the simultaneous interpolation of the data together with the continuum extrapolation, the reduction of contributions from localized statistical fluctuations as well as overcoming the difficulty associated with having access to different physical magnetic fields depending on the volume of each lattice.

\bibliography{biblio}% Produces the bibliography via BibTeX.

\providecommand{\noopsort}[1]{}\providecommand{\singleletter}[1]{#1}%
\begin{thebibliography}{37}%
\makeatletter
\providecommand \@ifxundefined [1]{%
 \@ifx{#1\undefined}
}%
\providecommand \@ifnum [1]{%
 \ifnum #1\expandafter \@firstoftwo
 \else \expandafter \@secondoftwo
 \fi
}%
\providecommand \@ifx [1]{%
 \ifx #1\expandafter \@firstoftwo
 \else \expandafter \@secondoftwo
 \fi
}%
\providecommand \natexlab [1]{#1}%
\providecommand \enquote  [1]{``#1''}%
\providecommand \bibnamefont  [1]{#1}%
\providecommand \bibfnamefont [1]{#1}%
\providecommand \citenamefont [1]{#1}%
\providecommand \href@noop [0]{\@secondoftwo}%
\providecommand \href [0]{\begingroup \@sanitize@url \@href}%
\providecommand \@href[1]{\@@startlink{#1}\@@href}%
\providecommand \@@href[1]{\endgroup#1\@@endlink}%
\providecommand \@sanitize@url [0]{\catcode `\\12\catcode `\$12\catcode `\&12\catcode `\#12\catcode `\^12\catcode `\_12\catcode `\%12\relax}%
\providecommand \@@startlink[1]{}%
\providecommand \@@endlink[0]{}%
\providecommand \url  [0]{\begingroup\@sanitize@url \@url }%
\providecommand \@url [1]{\endgroup\@href {#1}{\urlprefix }}%
\providecommand \urlprefix  [0]{URL }%
\providecommand \Eprint [0]{\href }%
\providecommand \doibase [0]{http://dx.doi.org/}%
\providecommand \selectlanguage [0]{\@gobble}%
\providecommand \bibinfo  [0]{\@secondoftwo}%
\providecommand \bibfield  [0]{\@secondoftwo}%
\providecommand \translation [1]{[#1]}%
\providecommand \BibitemOpen [0]{}%
\providecommand \bibitemStop [0]{}%
\providecommand \bibitemNoStop [0]{.\EOS\space}%
\providecommand \EOS [0]{\spacefactor3000\relax}%
\providecommand \BibitemShut  [1]{\csname bibitem#1\endcsname}%
\let\auto@bib@innerbib\@empty
%</preamble>
\bibitem [{\citenamefont {Kharzeev}\ \emph {et~al.}(2024)\citenamefont {Kharzeev}, \citenamefont {Liao},\ and\ \citenamefont {Tribedy}}]{Kharzeev:2024zzm}%
  \BibitemOpen
  \bibfield  {author} {\bibinfo {author} {\bibfnamefont {D.~E.}\ \bibnamefont {Kharzeev}}, \bibinfo {author} {\bibfnamefont {J.}~\bibnamefont {Liao}}, \ and\ \bibinfo {author} {\bibfnamefont {P.}~\bibnamefont {Tribedy}},\ }\href {\doibase 10.1142/9789811294679_0006} {\bibfield  {journal} {\bibinfo  {journal} {Int. J. Mod. Phys. E}\ }\textbf {\bibinfo {volume} {33}},\ \bibinfo {pages} {2430007} (\bibinfo {year} {2024})},\ \Eprint {http://arxiv.org/abs/2405.05427} {arXiv:2405.05427 [nucl-th]} \BibitemShut {NoStop}%
\bibitem [{\citenamefont {Fukushima}\ \emph {et~al.}(2008)\citenamefont {Fukushima}, \citenamefont {Kharzeev},\ and\ \citenamefont {Warringa}}]{Fukushima:2008xe}%
  \BibitemOpen
  \bibfield  {author} {\bibinfo {author} {\bibfnamefont {K.}~\bibnamefont {Fukushima}}, \bibinfo {author} {\bibfnamefont {D.~E.}\ \bibnamefont {Kharzeev}}, \ and\ \bibinfo {author} {\bibfnamefont {H.~J.}\ \bibnamefont {Warringa}},\ }\href {\doibase 10.1103/PhysRevD.78.074033} {\bibfield  {journal} {\bibinfo  {journal} {Phys. Rev. D}\ }\textbf {\bibinfo {volume} {78}},\ \bibinfo {pages} {074033} (\bibinfo {year} {2008})},\ \Eprint {http://arxiv.org/abs/0808.3382} {arXiv:0808.3382 [hep-ph]} \BibitemShut {NoStop}%
\bibitem [{\citenamefont {Li}\ \emph {et~al.}(2016)\citenamefont {Li}, \citenamefont {Kharzeev}, \citenamefont {Zhang}, \citenamefont {Huang}, \citenamefont {Pletikosic}, \citenamefont {Fedorov}, \citenamefont {Zhong}, \citenamefont {Schneeloch}, \citenamefont {Gu},\ and\ \citenamefont {Valla}}]{Li:2014bha}%
  \BibitemOpen
  \bibfield  {author} {\bibinfo {author} {\bibfnamefont {Q.}~\bibnamefont {Li}}, \bibinfo {author} {\bibfnamefont {D.~E.}\ \bibnamefont {Kharzeev}}, \bibinfo {author} {\bibfnamefont {C.}~\bibnamefont {Zhang}}, \bibinfo {author} {\bibfnamefont {Y.}~\bibnamefont {Huang}}, \bibinfo {author} {\bibfnamefont {I.}~\bibnamefont {Pletikosic}}, \bibinfo {author} {\bibfnamefont {A.~V.}\ \bibnamefont {Fedorov}}, \bibinfo {author} {\bibfnamefont {R.~D.}\ \bibnamefont {Zhong}}, \bibinfo {author} {\bibfnamefont {J.~A.}\ \bibnamefont {Schneeloch}}, \bibinfo {author} {\bibfnamefont {G.~D.}\ \bibnamefont {Gu}}, \ and\ \bibinfo {author} {\bibfnamefont {T.}~\bibnamefont {Valla}},\ }\href {\doibase 10.1038/nphys3648} {\bibfield  {journal} {\bibinfo  {journal} {Nature Phys.}\ }\textbf {\bibinfo {volume} {12}},\ \bibinfo {pages} {550} (\bibinfo {year} {2016})},\ \Eprint {http://arxiv.org/abs/1412.6543} {arXiv:1412.6543 [cond-mat.str-el]} \BibitemShut {NoStop}%
\bibitem [{\citenamefont {Adamczyk}\ \emph {et~al.}(2013)\citenamefont {Adamczyk} \emph {et~al.}}]{STAR:2013ksd}%
  \BibitemOpen
  \bibfield  {author} {\bibinfo {author} {\bibfnamefont {L.}~\bibnamefont {Adamczyk}} \emph {et~al.} (\bibinfo {collaboration} {STAR}),\ }\href {\doibase 10.1103/PhysRevC.88.064911} {\bibfield  {journal} {\bibinfo  {journal} {Phys. Rev. C}\ }\textbf {\bibinfo {volume} {88}},\ \bibinfo {pages} {064911} (\bibinfo {year} {2013})},\ \Eprint {http://arxiv.org/abs/1302.3802} {arXiv:1302.3802 [nucl-ex]} \BibitemShut {NoStop}%
\bibitem [{\citenamefont {Adamczyk}\ \emph {et~al.}(2014)\citenamefont {Adamczyk} \emph {et~al.}}]{STAR:2014uiw}%
  \BibitemOpen
  \bibfield  {author} {\bibinfo {author} {\bibfnamefont {L.}~\bibnamefont {Adamczyk}} \emph {et~al.} (\bibinfo {collaboration} {STAR}),\ }\href {\doibase 10.1103/PhysRevLett.113.052302} {\bibfield  {journal} {\bibinfo  {journal} {Phys. Rev. Lett.}\ }\textbf {\bibinfo {volume} {113}},\ \bibinfo {pages} {052302} (\bibinfo {year} {2014})},\ \Eprint {http://arxiv.org/abs/1404.1433} {arXiv:1404.1433 [nucl-ex]} \BibitemShut {NoStop}%
\bibitem [{\citenamefont {Abdallah}\ \emph {et~al.}(2022)\citenamefont {Abdallah} \emph {et~al.}}]{STAR:2021mii}%
  \BibitemOpen
  \bibfield  {author} {\bibinfo {author} {\bibfnamefont {M.}~\bibnamefont {Abdallah}} \emph {et~al.} (\bibinfo {collaboration} {STAR}),\ }\href {\doibase 10.1103/PhysRevC.105.014901} {\bibfield  {journal} {\bibinfo  {journal} {Phys. Rev. C}\ }\textbf {\bibinfo {volume} {105}},\ \bibinfo {pages} {014901} (\bibinfo {year} {2022})},\ \Eprint {http://arxiv.org/abs/2109.00131} {arXiv:2109.00131 [nucl-ex]} \BibitemShut {NoStop}%
\bibitem [{\citenamefont {Yamamoto}(2015)}]{Yamamoto:2015fxa}%
  \BibitemOpen
  \bibfield  {author} {\bibinfo {author} {\bibfnamefont {N.}~\bibnamefont {Yamamoto}},\ }\href {\doibase 10.1103/PhysRevD.92.085011} {\bibfield  {journal} {\bibinfo  {journal} {Phys. Rev. D}\ }\textbf {\bibinfo {volume} {92}},\ \bibinfo {pages} {085011} (\bibinfo {year} {2015})},\ \Eprint {http://arxiv.org/abs/1502.01547} {arXiv:1502.01547 [cond-mat.mes-hall]} \BibitemShut {NoStop}%
\bibitem [{\citenamefont {Hou}\ \emph {et~al.}(2011)\citenamefont {Hou}, \citenamefont {Liu},\ and\ \citenamefont {Ren}}]{Hou:2011ze}%
  \BibitemOpen
  \bibfield  {author} {\bibinfo {author} {\bibfnamefont {D.}~\bibnamefont {Hou}}, \bibinfo {author} {\bibfnamefont {H.}~\bibnamefont {Liu}}, \ and\ \bibinfo {author} {\bibfnamefont {H.-c.}\ \bibnamefont {Ren}},\ }\href {\doibase 10.1007/JHEP05(2011)046} {\bibfield  {journal} {\bibinfo  {journal} {JHEP}\ }\textbf {\bibinfo {volume} {05}},\ \bibinfo {pages} {046} (\bibinfo {year} {2011})},\ \Eprint {http://arxiv.org/abs/1103.2035} {arXiv:1103.2035 [hep-ph]} \BibitemShut {NoStop}%
\bibitem [{\citenamefont {Buividovich}(2014)}]{Buividovich:2013hza}%
  \BibitemOpen
  \bibfield  {author} {\bibinfo {author} {\bibfnamefont {P.~V.}\ \bibnamefont {Buividovich}},\ }\href {\doibase 10.1016/j.nuclphysa.2014.02.022} {\bibfield  {journal} {\bibinfo  {journal} {Nucl. Phys. A}\ }\textbf {\bibinfo {volume} {925}},\ \bibinfo {pages} {218} (\bibinfo {year} {2014})},\ \Eprint {http://arxiv.org/abs/1312.1843} {arXiv:1312.1843 [hep-lat]} \BibitemShut {NoStop}%
\bibitem [{\citenamefont {Zubkov}(2016)}]{Zubkov:2016tcp}%
  \BibitemOpen
  \bibfield  {author} {\bibinfo {author} {\bibfnamefont {M.~A.}\ \bibnamefont {Zubkov}},\ }\href {\doibase 10.1103/PhysRevD.93.105036} {\bibfield  {journal} {\bibinfo  {journal} {Phys. Rev. D}\ }\textbf {\bibinfo {volume} {93}},\ \bibinfo {pages} {105036} (\bibinfo {year} {2016})},\ \Eprint {http://arxiv.org/abs/1605.08724} {arXiv:1605.08724 [hep-ph]} \BibitemShut {NoStop}%
\bibitem [{\citenamefont {Buividovich}(2024)}]{Buividovich:2024bmu}%
  \BibitemOpen
  \bibfield  {author} {\bibinfo {author} {\bibfnamefont {P.~V.}\ \bibnamefont {Buividovich}},\ }\href {\doibase 10.1103/PhysRevD.110.094508} {\bibfield  {journal} {\bibinfo  {journal} {Phys. Rev. D}\ }\textbf {\bibinfo {volume} {110}},\ \bibinfo {pages} {094508} (\bibinfo {year} {2024})},\ \Eprint {http://arxiv.org/abs/2404.14263} {arXiv:2404.14263 [hep-lat]} \BibitemShut {NoStop}%
\bibitem [{\citenamefont {Horv\'ath}\ \emph {et~al.}(2020)\citenamefont {Horv\'ath}, \citenamefont {Hou}, \citenamefont {Liao},\ and\ \citenamefont {Ren}}]{Horvath:2019dvl}%
  \BibitemOpen
  \bibfield  {author} {\bibinfo {author} {\bibfnamefont {M.}~\bibnamefont {Horv\'ath}}, \bibinfo {author} {\bibfnamefont {D.}~\bibnamefont {Hou}}, \bibinfo {author} {\bibfnamefont {J.}~\bibnamefont {Liao}}, \ and\ \bibinfo {author} {\bibfnamefont {H.-c.}\ \bibnamefont {Ren}},\ }\href {\doibase 10.1103/PhysRevD.101.076026} {\bibfield  {journal} {\bibinfo  {journal} {Phys. Rev. D}\ }\textbf {\bibinfo {volume} {101}},\ \bibinfo {pages} {076026} (\bibinfo {year} {2020})},\ \Eprint {http://arxiv.org/abs/1911.00933} {arXiv:1911.00933 [hep-ph]} \BibitemShut {NoStop}%
\bibitem [{\citenamefont {Brandt}\ \emph {et~al.}(2024{\natexlab{a}})\citenamefont {Brandt}, \citenamefont {Endr\H{o}di}, \citenamefont {Garnacho-Velasco},\ and\ \citenamefont {Mark\'o}}]{Brandt:2024wlw}%
  \BibitemOpen
  \bibfield  {author} {\bibinfo {author} {\bibfnamefont {B.~B.}\ \bibnamefont {Brandt}}, \bibinfo {author} {\bibfnamefont {G.}~\bibnamefont {Endr\H{o}di}}, \bibinfo {author} {\bibfnamefont {E.}~\bibnamefont {Garnacho-Velasco}}, \ and\ \bibinfo {author} {\bibfnamefont {G.}~\bibnamefont {Mark\'o}},\ }\href {\doibase 10.1007/JHEP09(2024)092} {\bibfield  {journal} {\bibinfo  {journal} {JHEP}\ }\textbf {\bibinfo {volume} {09}},\ \bibinfo {pages} {092} (\bibinfo {year} {2024}{\natexlab{a}})},\ \Eprint {http://arxiv.org/abs/2405.09484} {arXiv:2405.09484 [hep-lat]} \BibitemShut {NoStop}%
\bibitem [{\citenamefont {Deng}\ and\ \citenamefont {Huang}(2012)}]{Deng:2012pc}%
  \BibitemOpen
  \bibfield  {author} {\bibinfo {author} {\bibfnamefont {W.-T.}\ \bibnamefont {Deng}}\ and\ \bibinfo {author} {\bibfnamefont {X.-G.}\ \bibnamefont {Huang}},\ }\href {\doibase 10.1103/PhysRevC.85.044907} {\bibfield  {journal} {\bibinfo  {journal} {Phys. Rev. C}\ }\textbf {\bibinfo {volume} {85}},\ \bibinfo {pages} {044907} (\bibinfo {year} {2012})},\ \Eprint {http://arxiv.org/abs/1201.5108} {arXiv:1201.5108 [nucl-th]} \BibitemShut {NoStop}%
\bibitem [{\citenamefont {Brandt}\ \emph {et~al.}(2023{\natexlab{a}})\citenamefont {Brandt}, \citenamefont {Cuteri}, \citenamefont {Endr\H{o}di}, \citenamefont {Mark\'o}, \citenamefont {Sandbote},\ and\ \citenamefont {Valois}}]{Brandt:2023dir}%
  \BibitemOpen
  \bibfield  {author} {\bibinfo {author} {\bibfnamefont {B.~B.}\ \bibnamefont {Brandt}}, \bibinfo {author} {\bibfnamefont {F.}~\bibnamefont {Cuteri}}, \bibinfo {author} {\bibfnamefont {G.}~\bibnamefont {Endr\H{o}di}}, \bibinfo {author} {\bibfnamefont {G.}~\bibnamefont {Mark\'o}}, \bibinfo {author} {\bibfnamefont {L.}~\bibnamefont {Sandbote}}, \ and\ \bibinfo {author} {\bibfnamefont {A.~D.~M.}\ \bibnamefont {Valois}},\ }\href {\doibase 10.1007/JHEP11(2023)229} {\bibfield  {journal} {\bibinfo  {journal} {JHEP}\ }\textbf {\bibinfo {volume} {11}},\ \bibinfo {pages} {229} (\bibinfo {year} {2023}{\natexlab{a}})},\ \Eprint {http://arxiv.org/abs/2305.19029} {arXiv:2305.19029 [hep-lat]} \BibitemShut {NoStop}%
\bibitem [{\citenamefont {Buividovich}\ \emph {et~al.}(2009)\citenamefont {Buividovich}, \citenamefont {Chernodub}, \citenamefont {Luschevskaya},\ and\ \citenamefont {Polikarpov}}]{Buividovich:2009wi}%
  \BibitemOpen
  \bibfield  {author} {\bibinfo {author} {\bibfnamefont {P.~V.}\ \bibnamefont {Buividovich}}, \bibinfo {author} {\bibfnamefont {M.~N.}\ \bibnamefont {Chernodub}}, \bibinfo {author} {\bibfnamefont {E.~V.}\ \bibnamefont {Luschevskaya}}, \ and\ \bibinfo {author} {\bibfnamefont {M.~I.}\ \bibnamefont {Polikarpov}},\ }\href {\doibase 10.1103/PhysRevD.80.054503} {\bibfield  {journal} {\bibinfo  {journal} {Phys. Rev. D}\ }\textbf {\bibinfo {volume} {80}},\ \bibinfo {pages} {054503} (\bibinfo {year} {2009})},\ \Eprint {http://arxiv.org/abs/0907.0494} {arXiv:0907.0494 [hep-lat]} \BibitemShut {NoStop}%
\bibitem [{\citenamefont {Yamamoto}(2011{\natexlab{a}})}]{Yamamoto:2011gk}%
  \BibitemOpen
  \bibfield  {author} {\bibinfo {author} {\bibfnamefont {A.}~\bibnamefont {Yamamoto}},\ }\href {\doibase 10.1103/PhysRevLett.107.031601} {\bibfield  {journal} {\bibinfo  {journal} {Phys. Rev. Lett.}\ }\textbf {\bibinfo {volume} {107}},\ \bibinfo {pages} {031601} (\bibinfo {year} {2011}{\natexlab{a}})},\ \Eprint {http://arxiv.org/abs/1105.0385} {arXiv:1105.0385 [hep-lat]} \BibitemShut {NoStop}%
\bibitem [{\citenamefont {Yamamoto}(2011{\natexlab{b}})}]{Yamamoto:2011ks}%
  \BibitemOpen
  \bibfield  {author} {\bibinfo {author} {\bibfnamefont {A.}~\bibnamefont {Yamamoto}},\ }\href {\doibase 10.1103/PhysRevD.84.114504} {\bibfield  {journal} {\bibinfo  {journal} {Phys. Rev. D}\ }\textbf {\bibinfo {volume} {84}},\ \bibinfo {pages} {114504} (\bibinfo {year} {2011}{\natexlab{b}})},\ \Eprint {http://arxiv.org/abs/1111.4681} {arXiv:1111.4681 [hep-lat]} \BibitemShut {NoStop}%
\bibitem [{\citenamefont {Buividovich}\ \emph {et~al.}(2021)\citenamefont {Buividovich}, \citenamefont {Smith},\ and\ \citenamefont {von Smekal}}]{Buividovich:2020gnl}%
  \BibitemOpen
  \bibfield  {author} {\bibinfo {author} {\bibfnamefont {P.~V.}\ \bibnamefont {Buividovich}}, \bibinfo {author} {\bibfnamefont {D.}~\bibnamefont {Smith}}, \ and\ \bibinfo {author} {\bibfnamefont {L.}~\bibnamefont {von Smekal}},\ }\href {\doibase 10.1103/PhysRevD.104.014511} {\bibfield  {journal} {\bibinfo  {journal} {Phys. Rev. D}\ }\textbf {\bibinfo {volume} {104}},\ \bibinfo {pages} {014511} (\bibinfo {year} {2021})},\ \Eprint {http://arxiv.org/abs/2012.05184} {arXiv:2012.05184 [hep-lat]} \BibitemShut {NoStop}%
\bibitem [{\citenamefont {Puhr}\ and\ \citenamefont {Buividovich}(2018)}]{Puhr:2017ddx}%
  \BibitemOpen
  \bibfield  {author} {\bibinfo {author} {\bibfnamefont {M.}~\bibnamefont {Puhr}}\ and\ \bibinfo {author} {\bibfnamefont {P.~V.}\ \bibnamefont {Buividovich}},\ }\href {\doibase 10.1051/epjconf/201817504003} {\bibfield  {journal} {\bibinfo  {journal} {EPJ Web Conf.}\ }\textbf {\bibinfo {volume} {175}},\ \bibinfo {pages} {04003} (\bibinfo {year} {2018})},\ \Eprint {http://arxiv.org/abs/1712.01579} {arXiv:1712.01579 [hep-lat]} \BibitemShut {NoStop}%
\bibitem [{\citenamefont {Brandt}\ \emph {et~al.}(2024{\natexlab{b}})\citenamefont {Brandt}, \citenamefont {Endr\H{o}di}, \citenamefont {Garnacho-Velasco},\ and\ \citenamefont {Mark\'o}}]{Brandt:2023wgf}%
  \BibitemOpen
  \bibfield  {author} {\bibinfo {author} {\bibfnamefont {B.~B.}\ \bibnamefont {Brandt}}, \bibinfo {author} {\bibfnamefont {G.}~\bibnamefont {Endr\H{o}di}}, \bibinfo {author} {\bibfnamefont {E.}~\bibnamefont {Garnacho-Velasco}}, \ and\ \bibinfo {author} {\bibfnamefont {G.}~\bibnamefont {Mark\'o}},\ }\href {\doibase 10.1007/JHEP02(2024)142} {\bibfield  {journal} {\bibinfo  {journal} {JHEP}\ }\textbf {\bibinfo {volume} {02}},\ \bibinfo {pages} {142} (\bibinfo {year} {2024}{\natexlab{b}})},\ \Eprint {http://arxiv.org/abs/2312.02945} {arXiv:2312.02945 [hep-lat]} \BibitemShut {NoStop}%
\bibitem [{\citenamefont {Brandt}\ \emph {et~al.}(2024{\natexlab{c}})\citenamefont {Brandt}, \citenamefont {Endr\H{o}di}, \citenamefont {Mark\'o},\ and\ \citenamefont {Valois}}]{Brandt:2024blb}%
  \BibitemOpen
  \bibfield  {author} {\bibinfo {author} {\bibfnamefont {B.~B.}\ \bibnamefont {Brandt}}, \bibinfo {author} {\bibfnamefont {G.}~\bibnamefont {Endr\H{o}di}}, \bibinfo {author} {\bibfnamefont {G.}~\bibnamefont {Mark\'o}}, \ and\ \bibinfo {author} {\bibfnamefont {A.~D.~M.}\ \bibnamefont {Valois}},\ }\href {\doibase 10.1007/JHEP07(2024)027} {\bibfield  {journal} {\bibinfo  {journal} {JHEP}\ }\textbf {\bibinfo {volume} {07}},\ \bibinfo {pages} {027} (\bibinfo {year} {2024}{\natexlab{c}})},\ \Eprint {http://arxiv.org/abs/2405.06557} {arXiv:2405.06557 [hep-lat]} \BibitemShut {NoStop}%
\bibitem [{\citenamefont {Endr\H{o}di}(2025)}]{Endrodi:2024cqn}%
  \BibitemOpen
  \bibfield  {author} {\bibinfo {author} {\bibfnamefont {G.}~\bibnamefont {Endr\H{o}di}},\ }\href {\doibase 10.1016/j.ppnp.2024.104153} {\bibfield  {journal} {\bibinfo  {journal} {Prog. Part. Nucl. Phys.}\ }\textbf {\bibinfo {volume} {141}},\ \bibinfo {pages} {104153} (\bibinfo {year} {2025})},\ \Eprint {http://arxiv.org/abs/2406.19780} {arXiv:2406.19780 [hep-lat]} \BibitemShut {NoStop}%
\bibitem [{\citenamefont {Banerjee}\ \emph {et~al.}(2021)\citenamefont {Banerjee}, \citenamefont {Lewkowicz},\ and\ \citenamefont {Zubkov}}]{Banerjee:2021vvn}%
  \BibitemOpen
  \bibfield  {author} {\bibinfo {author} {\bibfnamefont {C.}~\bibnamefont {Banerjee}}, \bibinfo {author} {\bibfnamefont {M.}~\bibnamefont {Lewkowicz}}, \ and\ \bibinfo {author} {\bibfnamefont {M.~A.}\ \bibnamefont {Zubkov}},\ }\href {\doibase 10.1016/j.physletb.2021.136457} {\bibfield  {journal} {\bibinfo  {journal} {Phys. Lett. B}\ }\textbf {\bibinfo {volume} {819}},\ \bibinfo {pages} {136457} (\bibinfo {year} {2021})},\ \Eprint {http://arxiv.org/abs/2105.11391} {arXiv:2105.11391 [hep-ph]} \BibitemShut {NoStop}%
\bibitem [{\citenamefont {Dunne}(2004)}]{Dunne:2004nc}%
  \BibitemOpen
  \bibfield  {author} {\bibinfo {author} {\bibfnamefont {G.~V.}\ \bibnamefont {Dunne}},\ }\enquote {\bibinfo {title} {{Heisenberg-Euler effective Lagrangians: Basics and extensions}},}\ in\ \href {\doibase 10.1142/9789812775344_0014} {\emph {\bibinfo {booktitle} {{From fields to strings: Circumnavigating theoretical physics. Ian Kogan memorial collection (3 volume set)}}}},\ \bibinfo {editor} {edited by\ \bibinfo {editor} {\bibfnamefont {M.}~\bibnamefont {Shifman}}, \bibinfo {editor} {\bibfnamefont {A.}~\bibnamefont {Vainshtein}}, \ and\ \bibinfo {editor} {\bibfnamefont {J.}~\bibnamefont {Wheater}}}\ (\bibinfo {year} {2004})\ pp.\ \bibinfo {pages} {445--522},\ \Eprint {http://arxiv.org/abs/hep-th/0406216} {arXiv:hep-th/0406216} \BibitemShut {NoStop}%
\bibitem [{\citenamefont {Cao}(2018)}]{Cao:2017gqs}%
  \BibitemOpen
  \bibfield  {author} {\bibinfo {author} {\bibfnamefont {G.}~\bibnamefont {Cao}},\ }\href {\doibase 10.1103/PhysRevD.97.054021} {\bibfield  {journal} {\bibinfo  {journal} {Phys. Rev. D}\ }\textbf {\bibinfo {volume} {97}},\ \bibinfo {pages} {054021} (\bibinfo {year} {2018})},\ \Eprint {http://arxiv.org/abs/1801.00134} {arXiv:1801.00134 [nucl-th]} \BibitemShut {NoStop}%
\bibitem [{\citenamefont {Bors\'anyi}\ \emph {et~al.}(2010)\citenamefont {Bors\'anyi}, \citenamefont {Endr\H{o}di}, \citenamefont {Fodor}, \citenamefont {Jakov\'ac}, \citenamefont {Katz}, \citenamefont {Krieg}, \citenamefont {Ratti},\ and\ \citenamefont {Szab\'o}}]{Borsanyi:2010cj}%
  \BibitemOpen
  \bibfield  {author} {\bibinfo {author} {\bibfnamefont {S.}~\bibnamefont {Bors\'anyi}}, \bibinfo {author} {\bibfnamefont {G.}~\bibnamefont {Endr\H{o}di}}, \bibinfo {author} {\bibfnamefont {Z.}~\bibnamefont {Fodor}}, \bibinfo {author} {\bibfnamefont {A.}~\bibnamefont {Jakov\'ac}}, \bibinfo {author} {\bibfnamefont {S.~D.}\ \bibnamefont {Katz}}, \bibinfo {author} {\bibfnamefont {S.}~\bibnamefont {Krieg}}, \bibinfo {author} {\bibfnamefont {C.}~\bibnamefont {Ratti}}, \ and\ \bibinfo {author} {\bibfnamefont {K.~K.}\ \bibnamefont {Szab\'o}},\ }\href {\doibase 10.1007/JHEP11(2010)077} {\bibfield  {journal} {\bibinfo  {journal} {JHEP}\ }\textbf {\bibinfo {volume} {11}},\ \bibinfo {pages} {077} (\bibinfo {year} {2010})},\ \Eprint {http://arxiv.org/abs/1007.2580} {arXiv:1007.2580 [hep-lat]} \BibitemShut {NoStop}%
\bibitem [{\citenamefont {D{\"u}rr}(2013)}]{Durr:2013gp}%
  \BibitemOpen
  \bibfield  {author} {\bibinfo {author} {\bibfnamefont {S.}~\bibnamefont {D{\"u}rr}},\ }\href {\doibase 10.1103/PhysRevD.87.114501} {\bibfield  {journal} {\bibinfo  {journal} {Phys. Rev. D}\ }\textbf {\bibinfo {volume} {87}},\ \bibinfo {pages} {114501} (\bibinfo {year} {2013})},\ \Eprint {http://arxiv.org/abs/1302.0773} {arXiv:1302.0773 [hep-lat]} \BibitemShut {NoStop}%
\bibitem [{\citenamefont {Endr\H{o}di}(2013)}]{Endrodi:2013cs}%
  \BibitemOpen
  \bibfield  {author} {\bibinfo {author} {\bibfnamefont {G.}~\bibnamefont {Endr\H{o}di}},\ }\href {\doibase 10.1007/JHEP04(2013)023} {\bibfield  {journal} {\bibinfo  {journal} {JHEP}\ }\textbf {\bibinfo {volume} {04}},\ \bibinfo {pages} {023} (\bibinfo {year} {2013})},\ \Eprint {http://arxiv.org/abs/1301.1307} {arXiv:1301.1307 [hep-ph]} \BibitemShut {NoStop}%
\bibitem [{\citenamefont {Adhikari}\ and\ \citenamefont {Tiburzi}(2023)}]{Adhikari:2023fdl}%
  \BibitemOpen
  \bibfield  {author} {\bibinfo {author} {\bibfnamefont {P.}~\bibnamefont {Adhikari}}\ and\ \bibinfo {author} {\bibfnamefont {B.~C.}\ \bibnamefont {Tiburzi}},\ }\href {\doibase 10.1103/PhysRevD.107.094504} {\bibfield  {journal} {\bibinfo  {journal} {Phys. Rev. D}\ }\textbf {\bibinfo {volume} {107}},\ \bibinfo {pages} {094504} (\bibinfo {year} {2023})},\ \Eprint {http://arxiv.org/abs/2302.09179} {arXiv:2302.09179 [hep-lat]} \BibitemShut {NoStop}%
\bibitem [{\citenamefont {Brandt}\ \emph {et~al.}(2025)\citenamefont {Brandt}, \citenamefont {Endrodi}, \citenamefont {Garnacho~Velasco}, \citenamefont {Marko},\ and\ \citenamefont {Valois}}]{Brandt:2025now}%
  \BibitemOpen
  \bibfield  {author} {\bibinfo {author} {\bibfnamefont {B.~B.}\ \bibnamefont {Brandt}}, \bibinfo {author} {\bibfnamefont {G.}~\bibnamefont {Endrodi}}, \bibinfo {author} {\bibfnamefont {E.}~\bibnamefont {Garnacho~Velasco}}, \bibinfo {author} {\bibfnamefont {G.}~\bibnamefont {Marko}}, \ and\ \bibinfo {author} {\bibfnamefont {A.~D.~M.}\ \bibnamefont {Valois}},\ }\href {\doibase 10.22323/1.466.0196} {\bibfield  {journal} {\bibinfo  {journal} {PoS}\ }\textbf {\bibinfo {volume} {LATTICE2024}},\ \bibinfo {pages} {196} (\bibinfo {year} {2025})},\ \Eprint {http://arxiv.org/abs/2502.01155} {arXiv:2502.01155 [hep-lat]} \BibitemShut {NoStop}%
\bibitem [{\citenamefont {Itzykson}\ and\ \citenamefont {Zuber}(1980)}]{Itzykson:1980rh}%
  \BibitemOpen
  \bibfield  {author} {\bibinfo {author} {\bibfnamefont {C.}~\bibnamefont {Itzykson}}\ and\ \bibinfo {author} {\bibfnamefont {J.~B.}\ \bibnamefont {Zuber}},\ }\href@noop {} {\emph {\bibinfo {title} {{Quantum Field Theory}}}},\ International Series In Pure and Applied Physics\ (\bibinfo  {publisher} {McGraw-Hill},\ \bibinfo {address} {New York},\ \bibinfo {year} {1980})\BibitemShut {NoStop}%
\bibitem [{\citenamefont {Navas}\ \emph {et~al.}(2024)\citenamefont {Navas} \emph {et~al.}}]{ParticleDataGroup:2024cfk}%
  \BibitemOpen
  \bibfield  {author} {\bibinfo {author} {\bibfnamefont {S.}~\bibnamefont {Navas}} \emph {et~al.} (\bibinfo {collaboration} {Particle Data Group}),\ }\href {\doibase 10.1103/PhysRevD.110.030001} {\bibfield  {journal} {\bibinfo  {journal} {Phys. Rev. D}\ }\textbf {\bibinfo {volume} {110}},\ \bibinfo {pages} {030001} (\bibinfo {year} {2024})}\BibitemShut {NoStop}%
\bibitem [{\citenamefont {Endr\H{o}di}(2011)}]{endrHodi2011multidimensional}%
  \BibitemOpen
  \bibfield  {author} {\bibinfo {author} {\bibfnamefont {G.}~\bibnamefont {Endr\H{o}di}},\ }\href {\doibase 10.1016/j.cpc.2011.03.009} {\bibfield  {journal} {\bibinfo  {journal} {Comput. Phys. Commun.}\ }\textbf {\bibinfo {volume} {182}},\ \bibinfo {pages} {1307} (\bibinfo {year} {2011})},\ \Eprint {http://arxiv.org/abs/1010.2952} {arXiv:1010.2952 [physics.comp-ph]} \BibitemShut {NoStop}%
\bibitem [{\citenamefont {Brandt}\ \emph {et~al.}(2023{\natexlab{b}})\citenamefont {Brandt}, \citenamefont {Cuteri},\ and\ \citenamefont {Endr\H{o}di}}]{Brandt:2022hwy}%
  \BibitemOpen
  \bibfield  {author} {\bibinfo {author} {\bibfnamefont {B.~B.}\ \bibnamefont {Brandt}}, \bibinfo {author} {\bibfnamefont {F.}~\bibnamefont {Cuteri}}, \ and\ \bibinfo {author} {\bibfnamefont {G.}~\bibnamefont {Endr\H{o}di}},\ }\href {\doibase 10.1007/JHEP07(2023)055} {\bibfield  {journal} {\bibinfo  {journal} {JHEP}\ }\textbf {\bibinfo {volume} {07}},\ \bibinfo {pages} {055} (\bibinfo {year} {2023}{\natexlab{b}})},\ \Eprint {http://arxiv.org/abs/2212.14016} {arXiv:2212.14016 [hep-lat]} \BibitemShut {NoStop}%
\bibitem [{\citenamefont {Brandt}\ and\ \citenamefont {Endr\H{o}di}(2016)}]{Brandt:2016zdy}%
  \BibitemOpen
  \bibfield  {author} {\bibinfo {author} {\bibfnamefont {B.~B.}\ \bibnamefont {Brandt}}\ and\ \bibinfo {author} {\bibfnamefont {G.}~\bibnamefont {Endr\H{o}di}},\ }\href {\doibase 10.22323/1.256.0039} {\bibfield  {journal} {\bibinfo  {journal} {PoS}\ }\textbf {\bibinfo {volume} {LATTICE2016}},\ \bibinfo {pages} {039} (\bibinfo {year} {2016})},\ \Eprint {http://arxiv.org/abs/1611.06758} {arXiv:1611.06758 [hep-lat]} \BibitemShut {NoStop}%
\bibitem [{\citenamefont {Akaike}(1974)}]{Akaike:1974vps}%
  \BibitemOpen
  \bibfield  {author} {\bibinfo {author} {\bibfnamefont {H.}~\bibnamefont {Akaike}},\ }\href {\doibase 10.1109/TAC.1974.1100705} {\bibfield  {journal} {\bibinfo  {journal} {IEEE Trans. Automatic Control}\ }\textbf {\bibinfo {volume} {19}},\ \bibinfo {pages} {716} (\bibinfo {year} {1974})}\BibitemShut {NoStop}%
\end{thebibliography}%

\end{document}